\documentclass{LMCS}

\usepackage{proof}
\usepackage[all]{xy}
\usepackage{amstext}
\usepackage{amssymb}
\usepackage{latexsym}
\usepackage{hyperref}

\newcommand{\val}{\mathsf{val}}

\newcommand{\com}{\mathsf{com}}

\newcommand{\Spec}{\mathsf{Specs}}
\newcommand{\Assert}{\mathsf{Assertion}}

\newcommand{\new}{\mathsf{new}}
\newcommand{\free}{\mathsf{free}}

\newcommand{\ifz}{\mathsf{if}}
\newcommand{\fix}{\mathsf{fix}}
\newcommand{\mletin}[2]{\mathsf{let}\;{#1}\;\mathsf{in}\;{#2}}
\newcommand{\mlet}[1]{\mathsf{let}\;{#1}\;\mathsf{in}} 

\newcommand{\blank}{\mbox{-}}

\newcommand{\id}{\mathsf{id}}
\newcommand{\perm}{\mathsf{perm}}
\newcommand{\proj}{\mathsf{proj}}

\newcommand{\emp}{\mathsf{emp}}
\newcommand{\pointsto}{\mapsto}

\newcommand{\Loc}{\mathit{Loc}}
\newcommand{\Heap}{\mathit{Heap}}
\newcommand{\Obs}{\mathit{O}}
\newcommand{\Good}{\mathit{G}}

\newcommand{\cR}{\mathcal{R}}
\newcommand{\cont}{\mathit{Cont}}
\newcommand{\sval}{\mathit{Val}}
\newcommand{\sint}{\mathit{Int}}

\newcommand{\lfix}{\mathit{leastfix}}

\newcommand{\normal}{\mathit{normal}}
\newcommand{\error}{\mathit{err}}
\newcommand{\default}{\mathit{default}}

\newcommand{\mperp}{{\bot{\!\!}\bot}}
\newcommand{\fin}{\mathsf{fin}}
\newcommand{\FV}{\mathsf{fv}}
\newcommand{\dom}{\mathsf{dom}}
\newcommand{\bind}{{\scriptstyle\relax\rightarrow}}
\newcommand{\supp}{\mathsf{supp}}

\newcommand{\ff}[1]{[\![#1]\!]}

\newcommand{\key}[1]{{\sf #1}}

\newcommand{\ifthenElse}[3]{\key{if}\; #1 \;\key{then}\; #2 \;\key{else}\; #3}
\newcommand{\IfthenElse}[3]{\begin{array}[t]{@{}l}
                \key{if}\; #1\; \key{then}\; #2 \\
                \key{else}\; #3
                \end{array}}

\newcommand{\letin}[2]{\key{let}\; #1 \;\key{in}\; #2}

\newcommand{\Letin}[2]{\begin{array}[t]{@{}l}
                        \key{let}\; #1 \\
                        \key{in}\; #2
                       \end{array}}

\newcommand{\Letbe}[3]{\begin{array}[t]{@{}l}
                        \key{let}\; #1 \; \key{be}\; #2 \\
                        \key{in}\; #3
                       \end{array}}

\newcommand{\mtri}[3]{\{{#1}\}{#2}\{{#3}\}}
\newcommand{\ctri}[3]{\{{#1}\}\,{#2}\,\{{#3}\}}

\newcommand{\squad}[4]{{[{#1}]}
                        ({#2},  {#3})
                       {[{#4}]}}

\newcommand{\EQ}{\mathsf{eq}}

\newcommand{\defeq}{\stackrel{\mathit{def}}{=}}
\newcommand{\defsiff}{\stackrel{\mathit{def}}{\Leftrightarrow}}
\newcommand{\defliff}{\stackrel{\mathit{def}}{\iff}}

\newcommand{\cps}{\mathsf{cps}}

\newcommand{\init}{\mathsf{init}}
\newcommand{\inc}{\mathsf{inc}}
\newcommand{\mread}{\mathsf{read}}

\newcommand{\then}{\mathsf{then}}
\newcommand{\melse}{\mathsf{else}}

\newcommand{\data}{\mathsf{data}}
\newcommand{\mnext}{\mathsf{next}}

\def\doi{4 (2:6) 2008}
\lmcsheading%
{\doi}
{1--27}
{}
{}
{Sep.~21, 2007}
{May~15, 2008}
{}

\begin{document}

\title[Relational Parametricity and Separation Logic]
{Relational Parametricity and Separation Logic}

\author[L.~Birkedal]{Lars Birkedal\rsuper a}
\address{{\lsuper a}IT University of Copenhagen, Denmark}
\email{birkedal@itu.dk}

\author[H.~Yang]{Hongseok Yang\rsuper b}
\address{{\lsuper b}Queen Mary, University of London, UK}
\email{hyang@dcs.qmul.ac.uk}


\begin{abstract}
  Separation logic is a recent extension of Hoare logic for reasoning about
  programs with references to shared mutable data structures.  In this paper, we
  provide a new interpretation of the logic for a programming language with
  higher types. Our interpretation is based on Reynolds's relational
  parametricity, and it provides a formal connection between separation
  logic and data abstraction.
\end{abstract}

\maketitle

\section{Introduction}

Separation
logic~\cite{reynolds02,
ohearn-reynolds-yang01,birkedal-torpsmith-reynolds-popl04}
is a Hoare-style program logic, and variants of it have been applied
to prove correct interesting pointer algorithms such as copying a dag,
disposing a graph, the Schorr-Waite graph algorithm, and Cheney's
copying garbage collector.  The main advantage of separation logic
compared to ordinary Hoare logic is that it facilitates \emph{local
  reasoning}, formalized via the so-called \emph{frame rule} using a
connective called \emph{separating conjunction}.  The development of
separation logic initially focused on \emph{low-level} languages with
heaps and pointers, although in recent
work~\cite{yang-ohearn-reynolds-popl04,birkedal-torpsmith-yang-lics05} 
it was shown how to extend
separation logic first to
languages with a simple kind of
procedures~\cite{yang-ohearn-reynolds-popl04}
and then to languages also with
higher-types~\cite{birkedal-torpsmith-yang-lics05}.
Moreover, in~\cite{yang-ohearn-reynolds-popl04} a second-order frame rule
was proved sound and in~\cite{birkedal-torpsmith-yang-lics05} a whole range
of higher-order frame rules were proved sound for a separation-logic type
system. 

In~\cite{yang-ohearn-reynolds-popl04}
and~\cite{birkedal-torpsmith-yang-lics05}
it was explained how second and higher-order frame rules can be used
to reason about static imperative modules. The idea is roughly as
follows. Suppose that
we prove a specification for a client $c$, depending
on a module $k$,
$$
   \ctri{P_1}{k}{Q_1} 
   \;\vdash\; \ctri{P}{c(k)}{Q}.
$$
The proof of the client depends only on the ``abstract specification''
of the module, which describes the external behavior of $k$. 
Suppose further that an actual implementation 
$m$ of the module satisfies
$$
    \ctri{P_1 * I}{m}{Q_1 *I}.
$$
Here $I$ is the internal resource invariant of the module $m$, describing
the internal heap storage used by the module  to implement the abstract
specification. We can then employ a (higher-order) frame rule on the specification for the
client to get
$$
  \ctri{P_1 *I}{k}{Q_1*I}
  \;\vdash\; \ctri{P*I}{c(k)}{Q*I},
$$
and combine it with the specification for $m$ to obtain
$$
  \ctri{P*I}{c(m)}{Q*I}.
$$
A key advantage of this approach to modularity is that it facilitates
so-called ``ownership transfer.'' For example, if the module is a queue, 
then the ownership of cells transfers from the client to module upon
insertion into the queue. Moreover, the discipline allows clients to
maintain pointers into cells that have changed ownership to the module. 
See~\cite{yang-ohearn-reynolds-popl04} for examples and more explanations
of these facts.

Note that the higher-order frame rules in essence provide implicit
quantification over internal resource invariants.
In~\cite{biering-birkedal-torpsmith-esop05}
it is shown how one can employ a higher-order version of separation logic,
with explicit quantification of assertion predicates to reason about dynamic
modularity (where there can be several instances of the same abstract data
type implemented by an imperative module), see
also~\cite{parkinson-bierman-popl05}. The idea is to
existentially quantify over the internal resource invariants in
a module, so that in the above example,
$c$ would depend on a specification for $k$ of the form
$$
  \exists I.\; \ctri{P_1 * I}{k}{Q_1 *I}.
$$
As emphasized in the papers mentioned above, note that,
both in the case of implicit quantification over internal resource
invariants (higher-order frame rules) and in the case of explicit
quantification over internal resource invariants (existentials over
assertion predicates), reasoning about a client does not depend on the
internal resource invariant of possible module implementations.
Thus the methodology allows us to formally reason about \emph{mutable
abstract data types}, aka. \emph{imperative modules}.

However, the semantic models in the papers mentioned 
above do not allow us to make all the conclusions we would expect from
reasoning about mutable abstract data types. 
In particular, we would expect that \emph{clients should behave parametrically 
in the internal resource invariants}: When a client is applied to 
two different implementations of a mutable abstract data type, 
it should be the case that the client preserves 
relations between the internal resource invariants of the two
implementations. This is analogous to Reynolds's style 
relational parametricity for abstract data types with quantification over
type variables~\cite{reynolds}.

\begin{figure*}[t]
\hrule
$\,$

$$
\begin{array}{r@{\;\;}c@{\;\;}l@{\qquad}r@{\;\;}c@{\;\;}l@{\qquad}r@{\;\;}c@{\;\;}l}
\init_0(i) & \equiv & {c.\mnext := i}
&
\inc_0 & \equiv & 
\begin{array}[t]{@{}l@{}}
  \mlet{i = c.\mnext} \\
  \mlet{v = i.\data} \\
  {i.\data := v{+}1}
\end{array}
&
\mread_0 & \equiv &
\begin{array}[t]{@{}l@{}} 
\mlet{i = c.\mnext} \\
\mlet{v = i.\data} \\
{g.\data := v}
\end{array}
\\
\\
\init_1(i) & \equiv & 
\begin{array}[t]{@{}l@{}}
 \mlet{v = i.\data} \\
 {i.\data := {-}v;} \\
 {c.\mnext := i}
\end{array}
&
\inc_1 & \equiv & 
\begin{array}[t]{@{}l@{}}
  \mlet{i = c.\mnext} \\
  \mlet{v = i.\data} \\
  {i.\data := v{-}1}
\end{array}
&
\mread_1 & \equiv &
\begin{array}[t]{@{}l@{}} 
\mlet{i = c.\mnext} \\
\mlet{v = i.\data} \\
{g.\data := -v}
\\
\\
\end{array}
\end{array}
$$

\hrule
\caption{Counter Modules}
\label{fig:counters}
\end{figure*}

To understand this issue more clearly, 
consider the two implementations of a counter 
in Figure~\ref{fig:counters}.
A counter has three operations: 
$\init(i)$ for initializing the counter, 
and $\inc$ and $\mread$ for increasing and reading
the value of the counter.
In the first implementation, 
$\init_0(i)$
takes a heap cell $i$ containing
an initial value for the counter, and stores its address $i$ 
in the internal variable $c$, thereby setting the value
of the counter to the contents of $i$. The intention is that
when a client program calls this 
initialization routine with cell $i$, it should
transfer the ownership of the cell to the counter -- 
it should not dereference the cell after calling $\init_0(i)$.
The operation $\inc_0$ increases the value of 
the transferred cell $i$, and $\mread_0$ returns
the value of cell $i$, by storing it in a pre-determined
global variable $g$. The second 
implementation is almost identical to the first, except that
the value of the counter is negated. Thus, when
$R$ is the relation that relates a heap containing 
cell $i$ and variable $c$ with the same heap with the value of 
cell $i$
negated, all operations of these two implementations preserve
this relation $R$.

Now suppose that we are given a client program of the form
$$
\mlet{i{=}\new}
\bigl(i.\data{:=}n;\; \init(i);\; b(\inc,\mread)\bigr)
$$
whose body $b$ satisfies the following
specification in separation logic:
$$
  \ctri{\emp}{\inc}{\emp},\;\;\;
  \ctri{g\pointsto\blank}{\mread}{g\pointsto\blank}
  \,\;\vdash\,\;
  \ctri{P}{b(\inc,\mread)}{Q}
$$
for some $P,Q$ that do not mention cell $i$.
We expect that the body $b$ of the client preserves the relation $R$ of the
two implementations, and that the client cannot detect the difference
between the two. Our expectation is based on the specification for $b$,
which says that the triple $\ctri{P}{b(\inc,\mread)}{Q}$ can be proved in
separation logic, assuming only the ``abstract specification'' of the
$\inc$ and $\mread$ operations, where all the internal resources of the
module, such as cell $i$, are hidden. This provability should prevent $b$
from accessing the internal resources of a counter directly and thus
detecting the difference between the two implementations.  However, none of the
existing models of separation logic can justify our expectation on the
client program above.

In this paper we provide a new parametric model of separation 
logic, which
captures that clients behave parametrically in internal resource
invariants of mutable abstract data types. For instance,
our model shows that $b(\inc,\mread)$ preserves
the relation $R$, and thus it behaves
in the same way no matter whether we use the first or second
implementation of a counter. In the present paper,
we will focus on the implicit approach to quantification over
internal resource invariants via higher-order frame rules, since it is
technically simpler than the explicit approach.\footnote{The reason is that the
implicit quantification of separation logic uses quantification in a very
disciplined way so that the usual reading of assertions as sets of heaps
can be maintained; if we use quantification without any restrictions, 
as in \cite{Benton-CSL06}, it
appears that we cannot have the usual reading of assertions as sets of
heaps because, then, the rule of consequence is not sound.}

Our new model of separation logic is based on two novel ideas.
The first is to read specifications in separation logic
as relations
between two programs. For instance, in our model, the Hoare 
triple $\ctri{P}{b(\inc,\mread)}{Q}$ describes 
a relationship between two instantiations
$\ff{b(\inc,\mread)}_{\eta_0}$ and $\ff{b(\inc,\mread)}_{\eta_1}$ of
the client's body $b$ by environments $\eta_0$ and $\eta_1$.
Intuitively, environment $\eta_i$ defines an implementation
of module operations $\inc$ and 
$\mread$, so $\ff{b(\inc,\mread)}_{\eta_i}$ means 
$b$ is linked with
the implementation $\eta_i(\inc)$ and $\eta_i(\mread)$.
Note that when used with appropriate $\eta_0,\eta_1$
(i.e., $\eta_i$ that maps $\inc$ and $\mread$ to
the meaning of $\inc_i$ and $\mread_i$), 
the triple expresses how 
$b(\inc_0,\mread_0)$ 
is related to $b(\inc_1,\mread_1)$.

The second idea is to parameterize the interpretation by
relations on heaps.
Mathematically, this means that the interpretation uses
a Kripke structure that consists of relations on heaps. 
The relation parameter describes
how the internal resource invariants of
two modules are related, and it lets us express the preservation
of this relation by client programs. In our counter example,
an appropriate parameter is the relation $R$ above. When
the triple $\ctri{P}{b(\inc,\mread)}{Q}$ is interpreted with
$R$ (and $\eta_i$ corresponding to $\inc_i,\mread_i$), 
it says, in particular, that
$b(\inc_0,\mread_0)$ and $b(\inc_1,\mread_1)$
should preserve the relation $R$ between the internal resources
of the two implementations of a counter.

\subsection{Related Work}
\label{sec:related-work} 
Technically, it has proven to be a very non-trivial problem to define
a parametric model for separation logic. 
One of the main technical challenges in developing a relationally
parametric model of separation logic, even for a simple first-order
language, is that the standard models of separation logic allow the
identity of locations to be observed in the model.  This means in
particular that allocation of new heap cells is not parametric because the
identity of the location of the allocated cell can be observed in the
model.  (We made this observation in earlier unpublished joint work with
Noah Torp-Smith, see~\cite[Ch.~6]{torp-smith-phd-2005}.)

This problem of non-parametric memory allocation has also been noticed by
recent work on data refinement for heap storage, which exploits semantic
ideas from separation logic
\cite{mijajlovic-torpsmith-ohearn-fsttcs04,mijajlovic-yang-aplas05}.
However, the work on data refinement does not provide a satisfactory
solution.  Either it avoids the problem by assuming that clients do not
allocate cells~\cite{mijajlovic-torpsmith-ohearn-fsttcs04}, or its solution
has difficulties for handling higher-order procedures and formalizing
(observational) equivalences, not refinements, between two implementations
of a mutable abstract data type \cite{mijajlovic-yang-aplas05}.

Our solution to this challenge is to define a more refined semantics of the
programming language using FM domain theory, in the style of Benton and
Leperchey~\cite{benton}, in which one can name locations but not observe the
identity of locations because of the built-in use of permutation of
locations. 
Part of the trick of \emph{loc. cit.} is to define the semantics
in a continuation-passing style so that one can ensure that new locations
are suitably fresh with respect to the remainder of the computation. (See
Section~\ref{sec:semantics-pl} for more details.)  Benton and Leperchey
used the FM domain-theoretic model to reason about contextual equivalence
and here we extend the approach to give a semantics of separation logic in
a continuation-passing style. We relate this new interpretation to the
standard direct-style interpretation of separation logic via the so-called
observation closure $(-)^\mperp$ of a relation, see
Section~\ref{sec:quadruples}.

The other main technical challenge in developing a relationally parametric
model of separation logic for reasoning about mutable abstract data types
is to devise a model which validates a wide range
of higher-order frame rules.  Our solution to this challenge is to
define an intuitionistic interpretation of the specification
logic over a Kripke structure, whose ordering relation intuitively
captures the framing-in of resources.
Technically, the intuitionistic interpretation, in particular the
associated Kripke monotonicity, is used to validate a generalized frame rule.
Further, to show that the semantics of the logic does indeed satisfy
Kripke monotonicity for the base case of triples, we
interpret triples using a universal quantifier, which
intuitively quantifies over resources that can possibly be framed in.
In the earlier non-parametric model of higher-order frame rules for
separation-logic typing in~\cite{birkedal-torpsmith-yang-lics05} we also
made use of a Kripke structure. The difference is that in the present work
the elements of the Kripke structure are \emph{relations} on heaps rather than
predicates on heaps because we build a \emph{relationally} parametric
model.

In earlier work, Banerjee and Naumann~\cite{banerjee:naumann:jacm} studied
relational parametricity for dynamically allocated heap objects in a
Java-like language. Banerjee and Naumann made use of a non-trivial semantic
notion of confinement to describe internal resources of a module; here
instead we use separation logic, in particular separating conjunction and
frame rules, to describe which resources are internal to the module.  Our
model directly captures that whenever a client has been proved correct in
separation logic with respect to an abstract view of a module, then it does
not matter how the module has been implemented internally. And, this holds
for a higher-order language with higher-order frame rules.

This paper is organized as follows.
In Section~\ref{sec:programs-and-assertions} we describe the
programming and assertion languages we consider and in
Section~\ref{sec:separation-logic} we define our version of separation
logic.  In Section~\ref{sec:semantics-pl} we define the semantics of our
programming language in the category of FM-cpos, and describe our
relational interpretation of separation logic in
Section~\ref{sec:reln-int-of-sep-logic}. In
Section~\ref{sec:general-construction} we present a general abstract
construction that provides models of specification logic with higher-order
frame rules and show that the semantics of the previous section is in fact
a special case of the general construction. 
Section~\ref{sec:quadruples}
relates our relational interpretation to the standard interpretation of
separation logic, and in Section~\ref{sec:abstraction-theorem} we present
the abstraction theorem that our parametric model validates. 
We describe examples in Section~\ref{sec:examples}, and
finally we conclude and discuss future work in
Section~\ref{sec:discussion-futurework}.

An extended abstract of this paper was presented at the FOSSACS 2007
conference~\cite{birkedal-yang-fossacs07}. This paper includes proofs that
were missing in the conference version, and describes a general
mathematical construction that lies behind our parametric model of
separation logic. We also include a new example that illustrates the
subtleties of the problems and results.

\section{Programs and Assertions}
\label{sec:programs-and-assertions}
In this paper, we consider a 
higher-order language with immutable stack variables.
The types and terms of the language are defined as follows:
{
$$
\begin{array}{@{}rrcl@{}}
\mbox{Types} &
\tau & ::= & \com
       \mid \val \,{\rightarrow}\, \tau
       \mid \tau \,{\rightarrow}\, \tau 
\\
\mbox{Expressions} &
E & ::= & 
  i \mid 0 \mid 1 \mid -1 \mid E+E \mid E-E
\\
\mbox{Terms} &
M & ::= & x 
    \mid \lambda i.\,M  
    \mid M\; E
    \mid \lambda x\colon \tau.\,M  
    \mid M\, M \\
&& \mid & 
    \fix\,M  
    \mid \ifz\,(E{=}E)\,\then\,M\,\melse\,M 
    \mid M;M \\
&& \mid &
    \mletin{i{=}\new}{M}
    \mid \free(E)
    \mid \mletin{i{=}E.f}{M}
    \mid E.f{:=}E \;\;(f\in\{0,1\})
\end{array}$$}
The language separates expressions $E$ from terms $M$.
Expressions denote heap-independent values, which
are either the address of a heap cell or an integer.
Expressions are bound to 
{\em stack variables} $i,j$. On the other hand,
terms denote possibly heap-dependent computations,
 and  they
are bound to {\em identifiers} $x,y$. The syntax
of the language ensures that expressions always
terminate, while terms can diverge. The types are
used to classify terms only. $\com$ denotes commands, 
$\val \rightarrow \tau$ means functions
that take an expression parameter, and
$\tau \rightarrow \tau'$ denotes functions that
takes a term parameter. Note that to support
two different function types, the language
includes two kinds of abstraction and application,
one for expression parameters and the other for term 
parameters. We assume that term parameters are
passed by name, and expression parameters are
passed by value.

To simplify the presentation, we take a simple storage
model where each heap cell has only two fields $0$ and $1$.
Command 
$\mletin{i{=}\new}{M}$ allocates such a binary heap cell, binds the 
address of the cell to $i$, and runs $M$ under this binding.
The $f$'th field of this newly allocated cell at address $i$ is read by
$\mletin{j=i.f}{N}$ and updated by
$i.f:=E$. The cell $i$ is deallocated by $\free(i)$.

\begin{figure*}[t]
\hrule
$\,$

$$
\begin{array}{c}
\infer{
  \Delta, i \vdash i
}{}
\qquad
\infer{\Delta \vdash 0}{}
\qquad
\infer{
\Delta \vdash E_1 + E_2
}{
\Delta \vdash E_1 &
\Delta \vdash E_2}
\qquad
\infer{
\Delta \vdash E_1 - E_2
}{
\Delta \vdash E_1 &
\Delta \vdash E_2
}
\\
\\
\infer{
  \Delta \mid \Gamma,x : \tau \vdash x : \tau
}{}
\quad
\infer{
  \Delta \mid \Gamma \vdash 
  \lambda i.\, M : \val \rightarrow \tau
}{
  \Delta, i \mid \Gamma \vdash M : \tau
}
\quad
\infer{
  \Delta \mid \Gamma \vdash 
  M\;E : \tau
}{
  \Delta \mid \Gamma \vdash M : \val \rightarrow \tau
  &
  \Delta \vdash E
}
\\
\\
\infer{
  \Delta \mid \Gamma 
  \vdash \lambda x : \tau.\, M : \tau \rightarrow \tau'
}{ 
  \Delta \mid \Gamma, x : \tau 
   \vdash M : \tau'
}
\quad
\infer{
  \Delta \mid \Gamma \vdash M\;N : \tau
}{ 
  \Delta \mid \Gamma \vdash M : \tau' \rightarrow \tau
  &
  \Delta \mid \Gamma \vdash N : \tau'
}
\quad
\infer{
  \Delta \mid \Gamma \vdash \fix\;M : \tau
}{
  \Delta \mid \Gamma \vdash M : \tau \rightarrow \tau
}
\\
\\
\infer{
  \Delta \mid \Gamma \vdash \ifz\;(E{=}F)\,\then\,M\,\melse\,N : \com
}{
  \Delta \vdash E
  &
  \Delta \vdash F
  &
  \Delta \mid \Gamma \vdash M : \com
  &
  \Delta \mid \Gamma \vdash N : \com
}
\\
\\
\infer{
  \Delta \,|\, \Gamma \vdash M;N \,{:}\, \com
}{
  \Delta \,|\, \Gamma \vdash M \,{:}\, \com
  &
  \Delta \,|\, \Gamma \vdash N \,{:}\, \com
}
\quad
\infer{
  \Delta \,|\, \Gamma \vdash \mletin{i{=}\new}{M} \,{:}\, \com
}{
  \Delta,i \,|\, \Gamma \vdash M \,{:}\, \com
}
\quad
\infer{
  \Delta \,|\, \Gamma \vdash {\free(E)} \,{:}\, \com
}{
  \Delta \vdash E
}
\\
\\
\infer[f\in\{0,1\}]{
  \Delta \mid \Gamma \vdash \mletin{i{=}E.f}{M} : \com
}{
  \Delta,i \mid \Gamma \vdash M : \com
  &
  \Delta \vdash E
}
\quad
\infer[f \in \{0,1\}]{
  \Delta \mid \Gamma \vdash {E.f:=F} : \com
}{
  \Delta \vdash E 
  &
  \Delta \vdash F
}
\end{array}
$$
\hrule
\caption{Typing Rules for Expressions and Terms}
\label{fig:typing}
\end{figure*}

The language uses typing judgments of the form
$\Delta \vdash E (\,{:}\,\val)$ and
$\Delta \,|\, \Gamma \vdash M \,{:}\, \tau$,
where $\Delta$ is a finite set of stack variables
and $\Gamma$ is a standard type environment for identifiers $x$.
The typing rules for expressions and terms are shown in
Figure~\ref{fig:typing}.

We use the standard assertions from separation logic to describe properties
of the heap:\footnote{We omit separating implication $-\!\!*$
to simplify presentation.}
{
$$
P
   \;\;::=\;\;  E = E 
  \,\mid\, E \leq E
  \,\mid\, E \pointsto E,E 
  \,\mid\, \emp
  \,\mid\, P * P
  \,\mid\, P \wedge P
  \,\mid\, \neg P
  \,\mid\, \exists i.\, P.
$$
}
The points-to predicate $E \pointsto E_0,E_1$ means
that the current heap has only one cell at address $E$ and 
that the $i$-th field of the cell has the value $E_i$.
The $\emp$ predicate denotes the empty heap, and the separating
conjunction $P*Q$ means that the current heap can be split into
two parts so that $P$ holds for the one and $Q$ holds for the other.
The other connectives have the usual meaning from classical logic.
All the missing connectives from classical logic are defined as usual. 

In the paper, we will use the three abbreviations 
$(E \pointsto \blank)$, $(E \pointsto \blank, E_1)$
and $(E \pointsto E_0,\blank)$. The first
$E \pointsto \blank$ is a syntactic sugar for 
$\exists i,j.\; E \pointsto i,j$, and denotes heaps
with cell $E$ only.
$E\pointsto \blank,E_1$ is an abbreviation for
$\exists i.\,E\pointsto i,E_1$, and means
 heaps that contain only cell $E$ and
store $E'$ in the second field of this unique cell $E$.
The last $E\pointsto E_0,\blank$ is defined similarly.

Assertions only depend on stack variables $i,j$, not identifiers $x,y$.
Thus assertions are typed by a judgment $\Delta \vdash P :
\Assert$.  The typing rules for this judgment are completely standard, and
thus omitted from this paper.

\section{Separation Logic}
\label{sec:separation-logic}
Our version of separation logic is the first-order 
{\em intuitionistic\/} logic extended with Hoare triples 
and invariant extension.
The formulas in the logic are called {\em specifications},
and they are defined by the following grammar:
{
$$
\begin{array}{rcl}
\varphi
  & ::= & \mtri{P}{M}{Q}
     \ | \ \varphi \otimes P
     \ | \ E = E
     \ | \ M = M 
\\
  & \ | \ & \varphi \wedge \varphi
    \ | \   \varphi \vee \varphi
    \ | \   \varphi \Rightarrow \varphi
    \ | \   \forall x\colon \tau. \varphi 
    \ | \   \exists x\colon \tau. \varphi
    \ | \   \forall i. \varphi 
    \ | \   \exists i. \varphi
\end{array}
$$
}
The formula $\varphi \otimes P$ means
the extension of $\varphi$ by the invariant $P$.
It can be viewed as 
a syntactic transformation of $\varphi$ that 
inserts $P*-$ into the pre and post conditions of all triples 
in $\varphi$.  For instance, 
$(\mtri{P}{x}{Q} \Rightarrow \mtri{P'}{M(x)}{Q'})  \otimes P_0$ is 
equivalent to 
$\mtri{P*P_0}{x}{Q*P_0} \Rightarrow \mtri{P'*P_0}{M(x)}{Q'*P_0}$.
We write $\Spec$ for the set of all specifications.

Specifications are typed by the judgment 
$\Delta \mid \Gamma \vdash \varphi : \Spec$, where we
overloaded $\Spec$ to mean the type for specifications.

The logic includes all the usual proof rules from first-order
intuitionistic logic with equality, and a rule for fixed-point induction.
In addition, it contains proof rules from separation logic, and
\emph{higher-order frame rules}, expressed in terms of rules for invariant
introduction and distribution. Figure~\ref{fig:sl-rules} shows some of
these additional rules and a rule for fixed-point 
induction. In the figure, we often omit contexts $\Delta \mid
\Gamma$ for specifications and also conditions about typing.

\begin{figure*}[t]
\hrule
$\,$

\begin{center}
{\sc Proof Rules for Hoare Triples}
\end{center}

$$
\begin{array}{@{}rcl@{}}
  (\forall i. \mtri{P}{M}{Q}) 
  & \;\Rightarrow\; &
  \mtri{\exists i.\,P}{M}{\exists i.\,Q}
  \quad
  (\mbox{where $i\not\in \FV(M)$})
\\[0.5ex]
  (\mtri{P}{M}{Q} \wedge \mtri{P'}{M}{Q'})
  & \;\Rightarrow\; &
  \mtri{P \vee P'}{M}{Q \vee Q'}
\\[0.5ex]
  \mtri{P \wedge E{=}F}{M}{Q}
  \wedge
  \mtri{P \wedge E{\not=}F}{N}{Q}
  & \;\Rightarrow\; &
  \mtri{P}{\ifz\,(E{=}F)\,\then\,M\,\melse\,N}{Q}
\\[0.5ex]
  \mtri{P}{M}{P_0} 
  \wedge
  \mtri{P_0}{N}{Q}
  & \;\Rightarrow\; &
  \mtri{P}{M;N}{Q}
\\[0.5ex]
  (\forall i.\,\mtri{P*i\pointsto 0,0}{M}{Q})
  & \;\Rightarrow\; &
  \mtri{P}{\mletin{i{=}\new}{M}}{Q}
  \;\hfill
  (\mbox{where $i{\not\in} \FV(P,Q)$})
\\[0.5ex]
  (\forall i.\, \mtri{P*E\pointsto i,E_1}{M}{Q})
  & \;\Rightarrow\; &
  \mtri{\exists i.\,P*E\pointsto i,E_1}{\mletin{i{=}E.0}{M}}{Q}
\\
&&
\hfill
  (\mbox{where $i{\not\in} \FV(E,Q)$})
\\[0.5ex]
\multicolumn{3}{c}{
  \mtri{E\pointsto \blank}{\free(E)}{\emp}
}
\\[0.5ex]
\multicolumn{3}{c}{
  \mtri{E\pointsto \blank,E_1}{E.0 := F}{E\pointsto F,E_1}
}
\\[2.5ex]
\multicolumn{3}{c}{
\infer{
  \Delta \mid \Gamma \vdash
   \mtri{P'}{M}{Q'} 
   \Rightarrow
   \mtri{P}{M}{Q}
}{
  \mbox{
  $\ff{P}_\rho \subseteq \ff{P'}_\rho$ and
  $\ff{Q'}_\rho \subseteq \ff{Q}_\rho$
  for all $\rho \in \ff{\Delta}$}
}
\quad
}
\\
\\
\end{array}
$$

\begin{center}
{\sc Proof Rules for Invariant Extension $-\otimes P$}
\end{center}

$$
\begin{array}{@{}rclrcl@{}}
  \varphi & \;\Rightarrow\; & \varphi \otimes P
&
  \mtri{P}{M}{P'} \otimes Q 
  & \;\Leftrightarrow\; &
  \mtri{P*Q}{M}{P'*Q} 
\\[0.5ex]
  (E = F) \otimes Q 
  & \;\Leftrightarrow\; &
  E=F
&
  (M = N) \otimes Q 
  & \;\Leftrightarrow\; &
  (M=N)
\\[0.5ex]
  (\varphi \otimes P) \otimes Q 
  & \;\Leftrightarrow\; &
  \varphi \otimes (P*Q)
&
  (\varphi \oplus \psi)\otimes P 
  & \;\Leftrightarrow\; &
  (\varphi \otimes P) \oplus (\psi \otimes P)
\\
&&&
\multicolumn{3}{r}{
  \qquad (\mbox{where } \oplus \in \{\Rightarrow, \wedge, \vee\})
}
\\[0.5ex]
  (\kappa x\colon \tau.\, \varphi)\otimes P 
  & \;\Leftrightarrow\; &
  \kappa x\colon \tau.\, \varphi \otimes P
&
  (\kappa i.\, \varphi)\otimes P 
  & \;\Leftrightarrow\; &
  \kappa i.\, \varphi \otimes P
\\
\multicolumn{3}{r}{
  \qquad(\mbox{where } \kappa \in \{\forall,\exists\})
}
&
\multicolumn{3}{r}{
  \qquad(\mbox{where } \kappa \in \{\forall,\exists\} \mbox{ and } i \not\in \FV(P))
}
\\
\\
\end{array}
$$

\begin{center}
{\sc Rule for Fixed-Point Induction}
\end{center}

$$
\begin{array}{@{}c@{}}
C ::= [\;] 
   \,{\mid}\, \lambda i. C
   \,{\mid}\, C\, E 
   \,{\mid}\, \lambda x\colon \tau. C
   \,{\mid}\, C\, M 
   \,{\mid}\, \fix\, C
   \,{\mid}\, C;M 
\qquad
\gamma ::= \mtri{P}{C}{Q}
  \,{\mid}\, \gamma {\wedge} \gamma 
  \,{\mid}\, \forall x\colon\tau. \gamma
  \,{\mid}\, \forall i. \gamma
\\[0.5ex]
(\forall x.\; \gamma(x) \Rightarrow \gamma(M\;x))
\;\Rightarrow\; 
\gamma(\fix\;M)
\end{array}
$$
where $\gamma(N)$ is a capture-avoiding insertion of $N$
into the hole $[-]$ in $\gamma$.
\vspace{1mm}
\hrule
\caption{Sample Proof Rules}
\label{fig:sl-rules}
\end{figure*}

The rules for Hoare triples are the standard proof rules
of separation logic adapted to our language. Note that in 
the rule of consequence, we use the standard
semantics of assertions $P,P',Q,Q'$,
in order to express semantic implications between those assertions
(of course, standard logical derivability
$\Delta\mid P\vdash P'$ and
$\Delta\mid Q'\vdash Q$ are
sufficient conditions).
The rules for invariant extension 
formalize higher-order frame rules,
extending the idea in \cite{birkedal-torpsmith-yang-lics05}.  The generalized higher-order 
frame rule $\varphi \Rightarrow \varphi \otimes P$ adds
an invariant $P$ to specification $\varphi$, and the other
rules distribute this added invariant all the way down to
the triples. 
We just show one use of those rules that lead
to the second-order frame rule: 
{
$$
\infer{
   \Delta \mid \Gamma,x\colon \com \vdash
   \mtri{P*P_0}{x}{Q*P_0} \Rightarrow \mtri{P'*P_0}{M(x)}{Q'*P_0}
}{
  \infer{
      \Delta \mid \Gamma,x\colon \com \vdash
      \mtri{P}{x}{Q}\otimes P_0 
      \Rightarrow 
      \mtri{P'}{M(x)}{Q'} \otimes P_0
  }{
    \infer{
        \Delta \mid \Gamma,x\colon \com \vdash
        (\mtri{P}{x}{Q}
         \Rightarrow 
        \mtri{P'}{M(x)}{Q'}) \otimes P_0
    }{
        \Delta \mid \Gamma,x\colon \com \vdash
        \mtri{P}{x}{Q}
         \Rightarrow
        \mtri{P'}{M(x)}{Q'}
    }
  }
}
$$
}
The last rule is for fixed-point induction,
and it relies on the restriction that
a specification is of the form $\gamma(\fix\;M)$. 
The grammar for $\gamma$ guarantees
that $\gamma(x)$ defines an admissible predicate for $x$,
thus ensuring the soundness of fixed-point induction.
Moreover, it also guarantees that $\gamma(x)$ holds 
when $M$ means $\bot$, so allowing us to
omit a usual base case, ``$\gamma(\bot)$,'' from the rule. 

Note that the rules do \emph{not} include the so-called 
conjunction rule:
$$
  (\mtri{P}{M}{Q} \wedge \mtri{P'}{M}{Q'})\;\Rightarrow\;
  \mtri{P \wedge P'}{M}{Q \wedge Q'}
$$
The omission of this rule is crucial, since
our parametricity interpretation does not validate
the rule. 
We discuss the conjunction rule further
in Section~\ref{sec:discussion-futurework}.

\begin{exa}\label{exa:client-counter}
Recall the counter example from the introduction and
consider the following simple client
$$
    \mletin{i{=}\new}{\bigl(i.0:=5;\init(i);\inc;\mread\bigr)},
$$
whose body consists of $\inc;\mread$.
The client initializes the value of the counter to $5$,
increases the counter, and reads the value of the counter.

In our logic, we can prove that the body of the client satisfies:
$$
  \Delta \,\mid\, \Gamma 
  \;\;\vdash\;\; 
  \varphi
  \;\;\Rightarrow\;\;
  \mtri{g\pointsto\blank}{\inc;\mread}{g\pointsto\blank}
$$
where $\Delta$ is a set of stack variables containing
$g$ and $\Gamma,\varphi$ are defined by
$$
  \Gamma
  \defeq
  \{\inc:\com, \mread:\com\},
\quad
  \varphi 
  \defeq
  \mtri{\emp}{\inc}{\emp} 
  \wedge 
  \mtri{g\pointsto\blank}{\mread}{g\pointsto\blank}.
$$
Note that cell $i$, which is transferred to the counter by $\init(i)$,
does not appear in any assertion of the specification for
the client's body. This implies, 
correctly, that the client does not dereference
the transferred cell $i$, after calling $\init(i)$.

The formal proof of the specification of the body
uses the first-order frame rule, and it is given below:
$$
\infer[6]{
  \Delta\mid\Gamma \,\vdash\,\varphi
  \Rightarrow \mtri{g\pointsto\blank}{\inc;\mread}{g\pointsto\blank}
}{
  \infer[4]{
    \Delta\mid\Gamma \,\vdash\,
    \varphi \Rightarrow \mtri{g\pointsto\blank}{\inc}{g\pointsto\blank}
  }{
    \infer[3]{
      \Delta\mid\Gamma \,\vdash\,
      \varphi \Rightarrow \mtri{\emp*g\pointsto\blank}{\inc}{\emp*g\pointsto\blank}
    }{
      \infer[2]{
        \Delta\mid\Gamma \,\vdash\,
        \varphi \Rightarrow (\mtri{\emp}{\inc}{\emp} \otimes (g\pointsto\blank))
      }{
        \infer[1]{
          \Delta\mid\Gamma \,\vdash\,
          \varphi \Rightarrow \mtri{\emp}{\inc}{\emp}
        }{}
      }
    }
  }
  &
  \infer[5]{
     \Delta \mid \Gamma \,\vdash\, \varphi \Rightarrow \mtri{g\pointsto\blank}{\mread}{g\pointsto\blank}
  }{}
}  
$$
The interesting parts of the proof are steps $2,3$,
where we use rules for invariant extensions, in order
to add the frame axiom $g\pointsto\blank$ into the pre
and post conditions of a triple. Note that
the addition of this frame axiom starts with a generalized
frame rule $\varphi \Rightarrow \varphi \otimes P$,
and continues with the rule that moves $P$ inside $\varphi$.
The remaining steps $1,5,4,6$ are instances of
 usual rules for first-order intuitionistic logic
or Hoare logic, such as the
elimination rule for conjunction
and the rule of Consequence.
\qed
\end{exa}

\section{Semantics of Programming Language}
\label{sec:semantics-pl}
Let $\Loc$ be a countably infinite set of locations.
The programming language is interpreted in the category 
of FM-cpos on $\Loc$.

We remind the reader of the basics of FM domain theory.
Call a bijection $\pi$ on $\Loc$ a {\em permutation\/}
when $\pi(l) \not= l$ only for finitely many $l$,
and let $\perm$ be the set of all permutations.
An FM-set is a pair of a set $A$ and a function
$\cdot$ of type $\perm \times A \rightarrow A$, such
that (1) $\id \cdot a  = a$ and 
$\pi \cdot (\pi' \cdot a) = (\pi \circ \pi') \cdot a$, and (2)
every $a \in A$ is {\em supported\/} 
by some finite subset $L$ of $\Loc$, i.e.,
$$
  \forall \pi \in \perm.\;
    (\forall l \in L.\; \pi(l) = l)
    \implies
    \pi \cdot a = a.
$$
It is known that every element $a$ in an FM-set $A$ has
a smallest set $L$ that supports $a$. This smallest set
is denoted $\supp(a)$. An FM function $f$ from an FM-set $A$ to an FM-set $B$ is
a function from $A$ to $B$ such that 
$f(\pi \cdot a) = \pi \cdot (f(a))$ for all $a,\pi$.

An FM-poset is an FM-set $A$ with a partial order $\sqsubseteq$ on 
$A$ such that $a \sqsubseteq b \implies \pi \cdot a \sqsubseteq \pi \cdot b$
for all $\pi,a,b$. We say that a ($\omega$-)chain $\{a_i\}_i$ in FM-poset
$A$ is {\em finitely supported\/} 
iff there is a finite subset $L$ of $\Loc$
that supports all elements in the chain. Finally, an 
FM-cpo is an FM-poset $(A,\sqsubseteq)$ for which
 every finitely-supported 
chain $\{a_i\}_i$ has a least upper bound, and an FM continuous
function $f$ from an FM-cpo $A$ to an FM-cpo $B$ is
an FM function from $A$ to $B$ that preserves the least upper bounds
of all finitely supported chains.

Types are interpreted as pointed FM-cpos,
using the categorical structure of the category of FM-cpos,
see Figure~\ref{fig:interpretation-types}. In the figure,
we use the FM-cpo $\sval$  of references defined by:
$$
\sval  \defeq \Loc + \sint + \{\default\}
$$
where $\pi \cdot v \defeq  \ifthenElse{(v \not\in \Loc)}{v}{\pi(v)}$ and $\default$ denotes a default value used for
type-incorrect expressions, such as the addition of two
locations.
The only nonstandard part
is the semantics of the command type
$\com$, which we define in the continuation passing
style following \cite{PittsAM:monsf,benton}:
$$
\begin{array}{@{}rclrcl@{}}
\Obs & \defeq & \{\normal,\error\}_\bot
\;
(\mbox{with } \pi \cdot o = o)
&
\Heap  & \defeq & 
\Loc \rightharpoonup_\fin \sval \times \sval
\\
\cont  & \defeq & (\Heap \rightarrow \Obs)
&
\ff{\com} & \defeq & (\Heap \times \cont \rightarrow \Obs).
\end{array}
$$
Here $A\times B$ and $A \rightarrow B$ are cartesian product and exponential
in the category of FM-cpos. And $A \rightharpoonup_\fin B$ is the FM-cpo
of the finite partial functions from $A$ to $B$ whose order
and permutation action are defined below:
\begin{enumerate}
\item $f \sqsubseteq g \defliff 
      \dom(f) = \dom(g) \;\mbox{and}\;  f(a) \sqsubseteq g(a) 
      \;\mbox{for all $a \in \dom(f)$,}$ 
\item $(\pi \cdot f)(a) \defeq \ifthenElse{\,(a \in \pi(\dom(f)))\,}{\,(\pi \cdot ((f \circ \pi^{-1})(a)))\,}{\,\mbox{undefined}}$.
\end{enumerate}

The first FM-cpo $\Obs$ specifies all 
possible observations, which are
normal termination $\normal$, erroneous termination $\error$
or divergence $\bot$. The next FM-cpo $\Heap$ 
denotes the set of heaps.  It formalizes that a heap
contains only finitely many allocated cells and each
cell in the heap has two fields. The third FM-cpo $\cont$
represents the set of continuations that consume heaps. Finally,
$\ff{\com}$ is the set of cps-style commands. Those commands
take a current heap $h$ and a continuation $k$,
and compute an observation in $O$ (often by 
computing a final heap $h'$, and
calling the given continuation $k$ with $h'$).

\begin{figure*}[t]
\hrule
$$
\begin{array}{rcl@{\qquad}rcl}
\sval & \defeq & \Loc + \sint + \{\default\}
& 
\Obs & \defeq &  \{\normal,\error\}_\bot
\\ 
\Heap & \defeq & \Loc \rightharpoonup_\fin \sval\times\sval
&
\cont & \defeq & \Heap \rightarrow \Obs
\\[1.5ex]
\ff{\val \rightarrow \tau} & \defeq & 
   \sval \rightarrow \ff{\tau}
&
\ff{\tau \rightarrow \tau'} & \defeq &
   \ff{\tau} \rightarrow \ff{\tau'}
\\
\ff{\com} & \defeq & \Heap \times \cont \rightarrow \Obs 
\\[1.5ex]
\ff{\Delta} & \defeq & \prod_{i \in \Delta} \sval
&
\ff{\Gamma} & \defeq & \prod_{x\colon\tau \in \Gamma} \ff{\tau}.
\end{array}
$$
\hrule
\caption{Interpretation of Types and Typing Contexts}
\label{fig:interpretation-types}
\end{figure*}

Note that $\Heap$ has the usual heap disjointness
predicate $h\#h'$, which denotes the disjointness of $\dom(h)$
and $\dom(h')$, and the usual partial heap combining operator
$\bullet$, which takes the union of (the graphs of) two
disjoint heaps. The $\#$ predicate and $\bullet$ operator fit
well with FM domain theory, because they 
preserve all permutations:
$h \# h' \iff (\pi \cdot h) \# (\pi \cdot h')$ and
$\pi \cdot (h \bullet h')  = 
 (\pi \cdot h) \bullet (\pi \cdot h')$.

The semantics of typing contexts $\Delta$ and $\Gamma$
is given by cartesian products:
$\ff{\Delta} \defeq \prod_{i\in\Delta} \sval$
and
$\ff{\Gamma} \defeq \prod_{x\colon\tau \in \Gamma} \ff{\tau}$.
The products here are taken over finite families,
so they give well-defined FM-cpos.\footnote{An infinite
product of FM-cpos is not necessarily an FM-cpo.} We
will use symbols $\rho$ and $\eta$ to denote environments
in $\ff{\Delta}$ and $\ff{\Gamma}$, respectively.

\begin{figure*}[t]
\hrule
$$
\begin{array}{c}
\ff{\Delta \vdash E}  \;\; : \;\;
 \ff{\Delta} \rightarrow \sval
\\[1.5ex]
\ff{\Delta,i \vdash i}_\rho \;\;\defeq\;\; \rho(i)
\qquad
\ff{\Delta \vdash 0}_\rho \;\;\defeq\;\; 0
\\
\ff{\Delta \vdash E_1+E_2}_\rho \;\;\defeq\;\;
\ifthenElse{(\ff{E_1}_\rho,\ff{E_2}_\rho \in \sint)}{(\ff{E_1}_\rho+\ff{E_2}_\rho)}{\default}
\\
\ff{\Delta \vdash E_1 - E_2}_\rho \;\; \defeq \;\;
\ifthenElse{(\ff{E_1}_\rho,\ff{E_2}_\rho \in \sint)}{(\ff{E_1}_\rho - \ff{E_2}_\rho)}{\default}
\end{array}
$$
\hrule
\caption{Interpretation of Expressions}
\label{fig:interpretation-expressions}
\end{figure*}

The semantics of expressions and terms is 
shown in Figures~\ref{fig:interpretation-expressions}
and \ref{fig:interpretation-terms}.
It is standard, except for the case of allocation, 
where we make use of the underlying FM domain theory:
The interpretation 
picks a location that is fresh with respect to
currently known values (i.e., $\supp(h,\eta,\rho)$)
as well as those that will be used by the continuation
(i.e., $\supp(k)$).  The cps-style
interpretation gives us an explicit handle on which
locations are used by the continuation, and
the FM domain theory ensures that
$\supp(h,\eta,\rho,k)$ is finite (so a new location $l$ can be chosen)
and that the choice of $l$ does not matter, as long as $l$ is not
in $\supp(h,\eta,\rho,k)$. (Formally, one shows by induction that
the semantics is well-defined.) We borrowed this interpretation from
Benton and Leperchey \cite{benton}.

\begin{figure*}[t]
\hrule
$$ 
\begin{array}{@{}r@{\,}c@{\,}l@{}}
\ff{\Delta \,|\, \Gamma \vdash M \colon \tau} & : & \ff{\Delta}\times\ff{\Gamma} \rightarrow \ff{\tau}
\\[1ex]
\ff{\Delta \,|\, \Gamma,x\colon\tau \vdash x \colon \tau}_{\rho,\eta} & \defeq & 
    \eta(x)
\\
\ff{\Delta \,|\, \Gamma \vdash \lambda i.\, M \colon \val\rightarrow \tau}_{\rho,\eta} & \defeq & 
\lambda v\colon \sval.\;
\ff{\Delta,i \,|\, \Gamma \vdash M \colon \tau}_{\rho[i\bind v],\eta} 
\\
\ff{\Delta \,|\, \Gamma \vdash M\;E \colon \tau}_{\rho,\eta} & \defeq & 
(\ff{\Delta \,|\, \Gamma \vdash M \colon \val \rightarrow \tau}_{\rho,\eta})\;
\ff{E}_\rho
\\
\ff{\Delta \,|\, \Gamma \vdash \lambda x\colon\tau'.\, M \colon \tau'\rightarrow \tau}_{\rho,\eta} & \defeq & 
\lambda m\colon \ff{\tau'}.\;
\ff{\Delta \,|\, \Gamma,x\colon\tau' \vdash M \colon \tau}_{\rho,\eta[x\bind m]} 
\\
\ff{\Delta \,|\, \Gamma \vdash M\;N \colon \tau}_{\rho,\eta} & \defeq & 
(\ff{\Delta \,|\, \Gamma \vdash M \colon \tau' \rightarrow \tau}_{\rho,\eta})\;
\ff{\Delta \,|\, \Gamma \vdash N \colon \tau'}_{\rho,\eta}
\\
\ff{\Delta \,|\, \Gamma \vdash \fix\;M \colon \tau}_{\rho,\eta} & \defeq & 
\lfix\;
\ff{\Delta \,|\, \Gamma \vdash M \colon \tau \rightarrow \tau}_{\rho,\eta}
\\
\ff{\Delta \,|\, \Gamma \vdash \ifz\,(E{=}F)\,\then\,M\,\melse\,N \colon \com}_{\rho,\eta} & \defeq & 
\IfthenElse{\ff{E}_\rho {=} \ff{F}_\rho}
  {\ff{\Delta \,|\, \Gamma \vdash M \colon \com}_{\rho,\eta}}
  {\ff{\Delta \,|\, \Gamma \vdash N \colon \com}_{\rho,\eta}}
\\
\ff{\Delta \,|\, \Gamma \vdash M;N \colon \com}_{\rho,\eta}(h,k) & \defeq & 
\Letbe
  {k'}
  {\lambda h'.\;\ff{\Delta \,|\, \Gamma \vdash N \colon \com}_{\rho,\eta}(h',k)}
  {\ff{\Delta \,|\, \Gamma \vdash M \colon \com}_{\rho,\eta}(h,k')}
\\
\ff{\Delta \,|\, \Gamma \,{\vdash}\, \mletin{i{=}\new}{M} \colon \com}_{\rho,\eta}(h,k) & \defeq &
  \ff{\Delta,i \,|\, \Gamma \vdash M \colon \com}_{\rho[i\bind l],\eta}
  (h\bullet [l\bind 0,0],k)
\\
  \multicolumn{3}{r}{
  \mbox{(where $l \in (\Loc{-}\supp(h,\rho,\eta,k))$)}
  }
\\
\ff{\Delta \,|\, \Gamma \,{\vdash}\, \free(E) \colon \com}_{\rho,\eta}(h,k) 
& \defeq &
  \IfthenElse
    {\ff{E}_\rho {\not\in} \dom(h)}
    {\error}
    {(k(h') \mbox{ for $h'$ s.t. $h'\bullet [\ff{E}_\rho\bind h(\ff{E}_\rho)] = h$})}
\\
\ff{\Delta \,|\, \Gamma \,{\vdash}\, \mletin{i{=}E.0}{M} \colon \com}_{\rho,\eta}(h,k) & \defeq &
  \IfthenElse
    {\ff{E}_\rho {\not\in} \dom(h)}
    {\error}
    {\Letin
       {(v,v') = h(\ff{E}_\rho)}
       {\ff{\Delta,i \,|\, \Gamma 
         \vdash M \colon \com}_{\rho[i\bind v],\eta}(h,k)}} 
\\
\ff{\Delta \,|\, \Gamma \vdash E.0 := F\colon \com}_{\rho,\eta}(h,k) & \defeq &
  \IfthenElse
    {\ff{E}_\rho {\not\in} \dom(h)}
    {\error}
    {(\letin
       {(v,v') = h(\ff{E}_\rho)}
       {k(h[\ff{E}_\rho \bind (\ff{F}_\rho,v')])})}
\end{array}
$$
\hrule
\caption{Interpretation of Terms}
\label{fig:interpretation-terms}
\end{figure*}

\section{Relational Interpretation of Separation Logic}
\label{sec:reln-int-of-sep-logic}
We now present the main result of this paper, 
a relational interpretation of separation logic.
In this interpretation, a specification
means a relation on terms,
rather than a set of terms
``satisfying'' the specification. 
This relational reading formalizes
the intuitive claim that proof rules in separation 
logic ensure parametricity with respect to the heap.

Our interpretation has two important components 
that ensure parametricity. The first is 
a Kripke structure $\cR$. The possible
worlds of $\cR$ are finitely supported binary 
relations $r$ on heaps,\footnote{A relation $r$ is 
finitely supported iff there is  
$L \subseteq_\fin \Loc$ s.t.
for every permutation $\pi$, if $\pi(l) = l$ for all 
$l\in L$,
then
$
   \forall h_0,h_1.\,
   h_0[r]h_1 \iff (\pi \cdot h_0)[r](\pi \cdot h_1). 
$} 
and the accessibility relation is
the preorder defined by the separating conjunction for relations: 
$$
\begin{array}{rcl}
   h_0[r * s]h_1  
   &
   \defsiff
   &
   \begin{array}[t]{@{}l@{}}
   \mbox{there exist splittings 
              $n_0\bullet m_0=h_0$ and $n_1\bullet m_1=h_1$
         such that} 
   \\
   \quad 
   n_0[r]n_1 \mbox{ and } m_0[s]m_1,
   \end{array}
\\
   r \sqsubseteq r' 
   & 
   \defsiff
   &
   \mbox{there exists $s$ such that $r*s = r'$}.
\end{array}
$$
Intuitively, $r \sqsubseteq r'$ means that $r'$
is a $*$-extension of $r$ by some $s$. The Kripke structure $\cR$ 
parameterizes our interpretation, and it guarantees 
that all the logical connectives behave parametrically wrt. relations
between internal resource invariants. 

The second is {\em semantic quadruples}, which describe
the relationship between two commands. We use
the semantic quadruples to interpret Hoare triples relationally.
Consider $c_0,c_1 \in \ff{\com}$
and $r,s \in \cR$. 
For each subset $D_0$ of an FM-cpo $D$,
define $\EQ(D_0)$ to be the partial identity relation on $D$ that 
equates only the elements in $D_0$. A {\em semantic quadruple\/} 
$\squad{r}{c_0}{c_1}{s}$ holds iff
$$
\begin{array}{l}
  \forall r' \in \cR.\,
  \forall h_0,h_1 \in \Heap.\,
  \forall k_0,k_1 \in \cont.\, 
\\
  \quad (h_0[r*r']h_1 \wedge k_0[s*r' \rightarrow \EQ(\Good)]k_1)
        \implies
        (c_0(h_0,k_0)[\EQ(\Good)]c_1(h_1,k_1)),
\end{array}
$$
where $\Good$ is the set $\Obs -\{\error\} = \{\normal,\bot\}$ of good
observations, and where $k_0[s*r' \rightarrow \EQ(\Good)]k_1$ means
that $k_0, k_1$ map heaps related in $s*r'$ into the diagonal of $\Good$.
The above condition 
indirectly expresses that if the input heaps $h_0,h_1$ are 
$r*r'$-related, then the output 
heaps are related by $s*r'$. Note that the definition 
quantifies over relations $r'$ for new heaps, thus implementing
relational parametricity.  In Section~\ref{sec:quadruples},
we show how semantic quadruples are related to a more
direct way of relating two commands and we also show
that the parametricity in the definition of semantic quadruples
implies the locality condition in separation logic~\cite{reynolds02}.

The semantics of the logic is defined by the satisfaction relation 
$\models_{\Delta|\Gamma}$ between
$\ff{\Delta} \times \ff{\Gamma}^2\times\cR$
and $\Spec$,
such that $\models_{\Delta|\Gamma}$ satisfies Kripke monotonicity:
$$
    (\rho,\eta_0,\eta_1, r \models_{\Delta|\Gamma} \varphi)  \,\wedge\, 
    (r \sqsubseteq r')
    \implies 
    (\rho,\eta_0,\eta_1, r' \models_{\Delta|\Gamma} \varphi).
$$ 
One way to understand the satisfaction relation is to
assume two machines that execute the same set of terms.
Each of these machines contains a chip that implements 
a module with a fixed set of operations. 
Intuitively, the
$(\rho,\eta_0,\eta_1,r)$ parameter of $\models$
specifies the configurations of those machines:
one machine uses $(\rho,\eta_0)$ to bind free stack variables
and identifiers of terms, and the other machine uses $(\rho,\eta_1)$
for the same purpose; and the internal resources of the
built-in modules
in those machines are related by $r$. The judgment $(\rho,\eta_0,\eta_1,r) \models_{\Delta|\Gamma} \varphi$
means that if two machines are configured by $(\rho,\eta_0,\eta_1,r)$,
then the meanings of the terms in two machines
are $\varphi$-related. Note that 
we allow different environments for the $\Gamma$ context only,
not for the $\Delta$ context. This is because we are mainly
concerned with parametricity with respect to the heap
and only $\Gamma$ entities, not $\Delta$ entities, 
depend on the heap. 


Figure~\ref{fig:definition-satisfaction} shows
the detailed interpretation of specifications. 
In the figure, we make use of the standard semantics of 
assertions~\cite{reynolds02}.
We now explain three cases in the definition of $\models$.

The first case is implication.
Our interpretation of implication exploits the
specific notion of accessibility in $\cR$.  It is
equivalent to the standard Kripke semantics of implication:
$$
\mbox{for all $r' \in \cR$,
if $r \sqsubseteq r'$ and $(\rho,\eta_0,\eta_1,r') \models \varphi$}, 
\mbox{ then $(\rho,\eta_0,\eta_1,r') \models \psi$},
$$
because $r \sqsubseteq r'$ iff $r' = r*s$ for some $s$.

The second case is quantification. If a stack
variable $i$ is quantified, we consider one semantic value, but
if an identifier $x$ is quantified, we consider two semantic 
values. This is again to reflect that in
our relational interpretation, we are mainly concerned with 
heap-dependent entities. Thus, we only read quantifiers for 
heap-dependent entities $x$ relationally.

The last case is invariant extension 
$\varphi\otimes P$. Mathematically, it says that if we extend the 
$r$ parameter by the partial equality for
predicate $P$, specification $\varphi$ holds. Intuitively,
this means that some heap cells not appearing in a specification $\varphi$ 
satisfy the invariant $P$.

\begin{figure*}[t]
\hrule
$$
\begin{array}{@{}l@{\,}c@{\,}l@{}}
\multicolumn{3}{l}{
\mbox{For all environments $\rho\in \ff{\Delta}$ 
      and $\eta_0,\eta_1 \in \ff{\Gamma}$ and all worlds $r \in \cR$,}
}
\\[1ex]
(\rho,\eta_0,\eta_1,r)
\models \mtri{P}{M}{Q}
& \defliff &
\squad{\EQ(\ff{P}_\rho)*r}{\ff{M}_{\rho,\eta_0}}{\ff{M}_{\rho,\eta_1}}{\EQ(\ff{Q}_\rho)*r}
\\
(\rho,\eta_0,\eta_1,r) 
\models \varphi \otimes P
& \defliff &
(\rho,\eta_0,\eta_1,r * \EQ(\ff{P}_\rho)) 
\models \varphi 
\\
(\rho,\eta_0,\eta_1,r) 
\models E = F
& \defliff &
\ff{E}_\rho = \ff{F}_\rho
\\
(\rho,\eta_0,\eta_1,r) 
\models M = N 
& \defliff &
\ff{M}_{\rho,\eta_0} = \ff{N}_{\rho,\eta_0}
\;\mbox{and}\;
\ff{M}_{\rho,\eta_1} = \ff{N}_{\rho,\eta_1}
\\
(\rho,\eta_0,\eta_1,r)
\models \varphi \Rightarrow \psi
& \defliff &
\begin{array}[t]{@{}l@{}}
\mbox{for all $s \in \cR$,
if $(\rho,\eta_0,\eta_1,r*s) \models \varphi$}, 
\\
\quad \mbox{then $(\rho,\eta_0,\eta_1,r*s) \models \psi$}
\end{array}
\\
(\rho,\eta_0,\eta_1,r)
\models \forall i.\,\varphi 
& \defliff &
\mbox{for all $v \in \sval$, }
(\rho[i\bind v],\eta_0,\eta_1,r)
\models \varphi 
\\
(\rho,\eta_0,\eta_1,r)
\models \exists i.\,\varphi 
& \defliff &
\mbox{there exists $v \in \sval$ s.t. }
(\rho[i\bind v],\eta_0,\eta_1,r)
\models \varphi 
\\
(\rho,\eta_0,\eta_1,r)
\models \forall x\colon\tau.\,\varphi 
& \defliff &
\mbox{for all $m,n \in \ff{\tau}$, }
(\rho,\eta_0[x\bind m],\eta_1[x\bind n],r)
\models \varphi 
\\
(\rho,\eta_0,\eta_1,r)
\models \exists x\colon\tau.\,\varphi 
& \defliff &
\mbox{there exist $m,n \in \ff{\tau}$ s.t. }
(\rho,\eta_0[x\bind m],\eta_1[x \bind n],r)
\models \varphi 
\\
(\rho,\eta_0,\eta_1,r)
\models \varphi \wedge \psi
& \defliff &
(\rho,\eta_0,\eta_1,r)
\models \varphi 
\;\mbox{and}\;
(\rho,\eta_0,\eta_1,r)
\models \psi
\\
(\rho,\eta_0,\eta_1,r)
\models \varphi \vee \psi
& \defliff &
(\rho,\eta_0,\eta_1,r)
\models \varphi 
\;\mbox{or}\;
(\rho,\eta_0,\eta_1,r)
\models \psi
\end{array}
$$
\hrule
\caption{Relational Interpretation of Separation Logic}
\label{fig:definition-satisfaction}
\end{figure*}

A specification $\Delta \mid \Gamma \vdash \varphi$ is \emph{valid}
iff $(\rho,\eta_0,\eta_1,r) \models \varphi$ holds
for all $(\rho,\eta_0,\eta_1,r)$.
A proof rule is \emph{sound}
when it is a valid axiom or an inference rule that concludes
a valid specification from valid premises.

\begin{lem}\label{lemma:axiom-extension}
The axioms for $\otimes$ are sound.
\end{lem}
\proof
All the axioms for $\otimes$ have the form
$\varphi \Rightarrow \psi$ or 
$\varphi \Leftrightarrow \psi$. When proving
those axioms, we use
the fact that $\varphi \Rightarrow \psi$ is valid
if and only if $(\rho,\eta_0,\eta_1,r) \models \varphi$
implies $(\rho,\eta_0,\eta_1,r) \models \psi$
for all $(\rho,\eta_0,\eta_1,r)$. 

First, consider the generalized frame rule 
$\varphi \Rightarrow \varphi \otimes P$. Suppose
that $(\rho,\eta_0,\eta_1,r) \models \varphi$. Then,
by Kripke monotonicity, 
$(\rho,\eta_0,\eta_1,r*\EQ(\ff{P}_\rho)) \models \varphi$.
Thus, $(\rho,\eta_0,\eta_1,r) \models \varphi \otimes P$.

Second, consider the distribution rule for triples.
We prove the validity of this rule as follows:
$$
\begin{array}{ll}
& (\rho,\eta_0,\eta_1,r) \models \mtri{P}{M}{Q} \otimes P_0
\\
\iff
&
(\rho,\eta_0,\eta_1,r*\EQ(\ff{P_0}_\rho)) \models \mtri{P}{M}{Q} 
\hfill
\quad(\mbox{by the semantics of $\otimes P$}).
\\
\iff
&  \squad
        {\EQ(\ff{P}_\rho)*\EQ(\ff{P_0}_\rho)*r}
        {\ff{M}_{\rho,\eta_0}}
        {\ff{M}_{\rho,\eta_1}}
        {\EQ(\ff{Q}_\rho)*\EQ(\ff{P_0}_\rho)*r}
\\
\iff
& \squad
        {\EQ(\ff{P*P_0}_\rho)*r}
        {\ff{M}_{\rho,\eta_0}}
        {\ff{M}_{\rho,\eta_1}}
        {\EQ(\ff{Q*P_0}_\rho)*r}
\\
\iff
& (\rho,\eta_0,\eta_1,r) \models \mtri{P*P_0}{M}{Q*P_0}
\hfill
\quad(\mbox{by the semantics of triples}).
\end{array}
$$
The second equivalence is by the semantics of triples,
and the third equivalence holds because $\EQ$ maps
$*$ for predicates to $*$ for relations.

Third, we prove the soundness of the distribution rules
for equality. Note that the semantics of $E=F$ and $M=N$
is independent of the heap relation $r$ in 
$(\rho,\eta_0,\eta_1,r)$. Thus, once we
fix the $\rho,\eta_0,\eta_1$ components, either
$E=F$ and $M=N$  hold for all $r$,
or $E=F$ and $M=N$  hold for no $r$.
Let $\varphi$ be $E=F$ or $M=N$. From the property of
$\varphi$ that we have just pointed out, it follows that
$$
  (\rho,\eta_0,\eta_1,r) \models \varphi
\iff
  (\rho,\eta_0,\eta_1,r*\EQ(\ff{P}_\rho)) \models \varphi
\iff
  (\rho,\eta_0,\eta_1,r) \models \varphi \otimes P.
$$

Finally, consider all the remaining rules, which
are distribution rules for logical connectives. All
cases can be proved
mostly by unrolling and rolling the definition of
$\models$. Here we explain two
cases. The first case is the distribution rule for 
existential quantification of $i$. We prove that this rule is 
sound below:
$$
\begin{array}{@{}ll@{}}
&
(\rho,\eta_0,\eta_1,r) \models \exists i.\,\varphi\otimes P
\\
\iff 
&
\mbox{there exists $v \in \val$ s.t. }
(\rho[i\bind v],\eta_0,\eta_1,r) \models \varphi\otimes P
\\
\iff 
&
\mbox{there exists $v \in \val$ s.t. }
(\rho[i\bind v],\eta_0,\eta_1,r*\EQ(\ff{P}_{\rho[i\bind v]})) 
\models \varphi
\\
\iff 
&
\mbox{there exists $v \in \val$ s.t. }
(\rho[i\bind v],\eta_0,\eta_1,r*\EQ(\ff{P}_{\rho}))
\models \varphi
\quad
\hfill
(\mbox{since $i \not\in \FV(P)$})
\\
\iff 
&
(\rho,\eta_0,\eta_1,r*\EQ(\ff{P}_{\rho}))
\models \exists i\colon \delta.\, \varphi
\\
\iff 
&
(\rho,\eta_0,\eta_1,r)
\models (\exists i\colon \delta.\, \varphi) \otimes P.
\end{array}
$$
All the equivalences except the third follow 
by rolling/unrolling the definition of $\models$.
The next case is the rule for implication,
which we prove sound as follows:
$$
\begin{array}{@{}ll@{}}
& (\rho,\eta_0,\eta_1,r) \models (\varphi \Rightarrow \psi) \otimes P
\\
\iff
& (\rho,\eta_0,\eta_1,r*\EQ(\ff{P}_\rho)) 
  \models \varphi \Rightarrow \psi
\\
\iff
& \forall s.\;
  \bigl((\rho,\eta_0,\eta_1,r*\EQ(\ff{P}_\rho)*s) \models \varphi\bigr)
  \implies
  \bigl((\rho,\eta_0,\eta_1,r*\EQ(\ff{P}_\rho)*s) \models \psi\bigr)
\\
\iff
& \forall s.\;
 \bigl((\rho,\eta_0,\eta_1,r*s) \models \varphi\otimes P\bigr)
 \implies
  \bigl((\rho,\eta_0,\eta_1,r*s) \models \psi\otimes P\bigr)
\\
\iff
& (\rho,\eta_0,\eta_1,r) \models 
  (\varphi\otimes P) \Rightarrow (\psi\otimes P).
\end{array}
$$
Again, all the equivalences are obtained by rolling/unrolling 
the definition of $\models$.
\qed

\begin{thm}\label{thm:soundness}
  All the proof rules are sound.
\end{thm}
\proof
The interpretation of all the logical connectives
is standard, so that the semantics validates
all the usual rules from first-order intuitionistic logic with 
equality. Moreover, by Lemma~\ref{lemma:axiom-extension}, 
all the rules about $\otimes$ are sound as well. Thus, 
it remains to show that the rules about Hoare triples
and fixed point induction are sound.

Note that most of the rules about triples and
fixed point induction have
the form $\varphi \Rightarrow \psi$. When proving
the soundness of those rules, we use
 the fact that  $\varphi \Rightarrow \psi$ is valid
if and only if $(\rho,\eta_0,\eta_1,r) \models \varphi$
implies $(\rho,\eta_0,\eta_1,r) \models \psi$
for all $(\rho,\eta_0,\eta_1,r)$. 
%
%
%
%

The first case is the rule for memory allocation:
$$
  (\forall i. \mtri{P*i\pointsto 0,0}{M}{Q})
  \;\Rightarrow\;
  \mtri{P}{\mletin{i{=}\new}{M}}{Q}.
$$
Consider $(\rho,\eta_0,\eta_1,r)$ satisfying the assumption
of the above axiom. We need to prove that 
$(\rho,\eta_0,\eta_1,r)$ also satisfies the conclusion, i.e.,
$$
   \squad
     {\EQ(\ff{P}_\rho)*r}
     {\ff{\mletin{i{=}\new}{M}}_{\rho,\eta_0}}
     {\ff{\mletin{i{=}\new}{M}}_{\rho,\eta_1}}
     {\EQ(\ff{Q}_\rho)*r}.
$$
Choose arbitrary $h_0,h_1 \in \Heap$, $k_0,k_1 \in \cont$,
and $s \in\cR$ such that 
$$
   h_0[\EQ(\ff{P}_\rho)*r*s]h_1
   \;\;\mbox{and}\;\;
   k_0[(\EQ(\ff{Q}_\rho)*r*s) \rightarrow \EQ(\Good)]k_1.
$$
Pick $l \in \Loc - \supp(h_0,h_1,\rho,\eta_0,\eta_1,k_0,k_1)$.
Then, the FM domain theory ensures that for $j=0,1$,
\begin{equation}\label{eqn:allocation}
  {\ff{\mletin{i{=}\new}{M}}_{\rho,\eta_j}}(h_j,k_j) 
  = 
  {\ff{M}_{\rho[i\bind l],\eta_j}}(h_j\bullet [l\bind 0,0],k_j).
\end{equation}
Let $\rho'$ be $\rho[i \bind l]$, and let $h'_j$ be 
$h_j\bullet [l\bind 0,0]$. We prove the required
relationship for $\mletin{i{=}\new}{M}$ as follows:
$$
\begin{array}{@{}ll@{}}
&
   h_0\bigl[\EQ(\ff{P}_\rho)*r*s\bigr]h_1\;\wedge\;
   k_0\bigl[(\EQ(\ff{Q}_\rho)*r*s) \rightarrow \EQ(\Good)\bigr]k_1
\\
\implies 
&
   h_0\bigl[\EQ(\ff{P}_{\rho'})*r*s\bigr]h_1
   \;\wedge\;
   k_0\bigl[(\EQ(\ff{Q}_{\rho'})*r*s) \rightarrow \EQ(\Good)\bigr]k_1
\\
\implies 
&
   h'_0\bigl[\EQ(\ff{P*i\pointsto 0,0}_{\rho'})*r*s\bigr]h'_1
   \;\wedge\;
   k_0\bigl[(\EQ(\ff{Q}_{\rho'})*r*s) \rightarrow \EQ(\Good)\bigr]k_1
\\
\implies 
&
  {\ff{M}}_{\rho',\eta_0}(h'_0,k_0) 
  \bigl[\EQ(\Good)\bigr]
  {\ff{M}}_{\rho',\eta_1}(h'_1,k_1)
\\
\implies
&
  {\ff{\mletin{i{=}\new}{M}}_{\rho,\eta_0}}(h_0,k_0) 
  \bigl[\EQ(\Good)\bigr]
  {\ff{\mletin{i{=}\new}{M}}_{\rho,\eta_1}}(h_1,k_1).
\end{array}
$$
The first implication holds, because 
$\rho$ and $\rho'$ are different only 
for $i$ but $i \not\in \FV(P,Q)$.
The second implication follows from the definition of $h'_j$,
and the third implication from the assumption that
$(\rho,\eta_0,\eta_1,r) 
     \models \forall i.\,\mtri{P*i\pointsto 0,0}{M}{Q}$.
Finally, the last implication holds, because
of the equation \ref{eqn:allocation}.

The second case is the axiom for lookup 
$$
(\forall i. \mtri{P * E\pointsto i,E_1}{M}{Q}) 
\;\Rightarrow\;
\mtri
  {\exists i. P * E\pointsto i,E_1}
  {\mletin{i{=}E.0}{M}}
  {Q}. 
$$
Consider $(\rho,\eta_0,\eta_1,r)$ that
satisfies $(\forall i. \mtri{P * E\pointsto i,E_1}{M}{Q})$,
and pick arbitrary $h_0,h_1 \in \Heap$,
$k_0,k_1 \in \cont$ and $s \in\cR$ such that
$$
 h_0\bigl[\EQ(\ff{\exists i. P*E\pointsto i,E_1}_\rho)*r*s
     \bigr]h_1
  \;\wedge\;
 k_0\bigl[\EQ(\ff{Q}_\rho)*r*s \rightarrow \EQ(\Good)\bigr]k_1.
$$
Let $l$ be $\ff{E}_\rho$ (which is well-defined since 
$i \not\in \FV(E)$).  By the first conjunct above, 
$l$ is in $\dom(h_0)\cap \dom(h_1)$, and
there exist $v$ and $\rho'$ such that 
$$
  v {=} \proj_0(h_0(l)) {=} \proj_0(h_1(l)),\;\;
  \rho'{=} \rho[i \bind v],\;\;\mbox{and}\;\;
  h_0\bigl[\EQ(\ff{P*E\pointsto i,E_1}_{\rho'})*r*s\bigr]h_1.
$$
Here $\proj_0$ is the projection of 
the first component of pairs.
The two equalities above about $v$ and $\rho'$ imply that
for $j = 0,1$, 
\begin{equation}\label{eqn:lookup}
\ff{\mletin{i{=}E.0}{M}}_{\rho,\eta_j}(h_j,k_j) 
= 
\ff{M}_{\rho',\eta_j}(h_j,k_j).
\end{equation}
We derive the desired relationship about $\mletin{i{=}E.0}{M}$
as follows:
$$
\begin{array}{@{}ll@{}}
&
  k_0\bigl[\EQ(\ff{Q}_\rho)*r*s \rightarrow \EQ(\Good)\bigr]k_1
  \;\wedge\;
  h_0\bigl[\EQ(\ff{P*E\pointsto i,E_1}_{\rho'})*r*s\bigr]h_1
\\
\implies &
  k_0\bigl[\EQ(\ff{Q}_{\rho'})*r*s \rightarrow \EQ(\Good)\bigr]k_1
  \;\wedge\;
  h_0\bigl[\EQ(\ff{P*E\pointsto i,E_1}_{\rho'})*r*s\bigr]h_1
\\
\implies &
  \ff{M}_{\rho',\eta_0}(h_0,k_0)
  \bigl[\EQ(\Good)\bigr]
  \ff{M}_{\rho',\eta_1}(h_1,k_1)
\\
\implies &
  \ff{\mletin{i{=}E.0}{M}}_{\rho,\eta_0}(h_0,k_0)
  \bigl[\EQ(\Good)\bigr]
  \ff{\mletin{i{=}E.0}{M}}_{\rho,\eta_1}(h_1,k_1).
\end{array}
$$
The first implication holds because $i \not\in \FV(Q)$,
the second follows from the fact that
$(\rho,\eta_0,\eta_1,r)$ satisfies the assumption of
this axiom, and the last implication follows from
the equation \ref{eqn:lookup}.

The third case is the axiom 
$\mtri{E\pointsto \blank}{\free(E)}{\emp}$. 
Choose arbitrary $(\rho,\eta_0,\eta_1,r)$,
$h_0,h_1 \in \Heap$, $k_0,k_1 \in \cont$, and
$s \in\cR$, such that
$$
   h_0\bigl[\EQ(\ff{E\pointsto \blank}_\rho) * r * s\bigr]h_1
   \;\wedge\;
   k_0\bigl[\EQ(\ff{\emp}_\rho) * r * s 
            \rightarrow \EQ(\Good)\bigr]k_1.
$$
By the first conjunct above,
there are splittings $m_0 \bullet n_0 = h_0$ and 
$m_1 \bullet n_1 = h_1$ such that $m_0[\EQ(\ff{E\pointsto \blank})]m_1$
and $n_0[r*s]n_1$. Note that the relationship between $m_0$ and $m_1$
implies that $\ff{\free(E)}_{\rho,\eta_j}(h_j,k_j) = k_j(n_j)$ for $j=0,1$.
Thus, it is sufficient to show that $k_0(n_0)[\EQ(\Good)]k_1(n_1)$.
Note that $n_0$ and $n_1$ are already related by $r*s$, 
and $k_0$ and $k_1$ by $\EQ(\ff{\emp}_\rho)*r*s \rightarrow \EQ(\Good)$.
The conclusion follows from these two relationships, because 
$\EQ({\ff{\emp}_\rho})*r*s = r*s$.

The fourth case is the axiom
$\mtri{E\pointsto \blank,E_1}{E.0:=F}{E \pointsto F,E_1}$.
Choose arbitrary $(\rho,\eta_0,\eta_1,r)$,
$h_0,h_1 \in \Heap$, $k_0,k_1 \in \cont$, and
$s \in\cR$, such that
$$
   h_0\bigl[\EQ(\ff{E\pointsto \blank,E_1}_\rho) * r * s\bigr]h_1
   \;\wedge\;
   k_0\bigl[\EQ(\ff{E\pointsto F,E_1}_\rho) * r * s 
            \rightarrow \EQ(\Good)\bigr]k_1.
$$
Because of the first conjunct, there
are splittings $m_0\bullet n_0 = h_0$ and 
$m_1 \bullet n_1 = h_1$ such that 
$m_0[\EQ(\ff{E \pointsto \blank,E_1}_\rho)]m_1$ and $n_0[r*s]n_1$.
Let $m'$ be the heap
$[\ff{E}_\rho \bind (\ff{F}_\rho,\ff{E_1}_\rho)]$. 
Then, we have the following two facts:
\begin{enumerate}
\item $(m' \bullet n_0) \bigl[\EQ(\ff{E\pointsto F,E_1}_\rho)*r*s\bigr] (m' \bullet n_1)$, and
\item for all $j = 0,1$, 
$\ff{E.0 := F}_{\rho,\eta_j}(h_j,k_j) 
  =
 k_j
 (m' \bullet n_j)$.
\end{enumerate}
By the first fact, $k_0(m'\bullet n_0)[\EQ(\Good)]k_1(m'\bullet n_1)$.
Now, the second fact gives the required
$\ff{E.0 := F}_{\rho,\eta_0}(h_0,k_0)\bigl[\EQ(\Good)\bigr]  
  \ff{E.0 := F}_{\rho,\eta_1}(h_1,k_1)$.

The fifth case is the rule of Consequence. Suppose
that $\ff{P}_\rho \subseteq \ff{P'}_\rho$ and
$\ff{Q'}_\rho \subseteq \ff{Q}_\rho$, and
$(\rho,\eta_0,\eta_1,r) \models \mtri{P'}{M}{Q'}$.
Consider
$h_0,h_1 \in \Heap$, $k_0,k_1 \in \cont$, and $s \in\cR$, such that
$$
  h_0[\EQ(\ff{P}_\rho)*r*s]h_1\;\;\wedge\;\;
  k_0[\EQ(\ff{Q}_\rho)*r*s \rightarrow \EQ(\Good)]k_1.
$$
Since $\EQ$ is monotone and $*$ preserves the subset order for
relations, 
$$
\begin{array}{rcl}
  \EQ(\ff{P}_\rho)*r*s & \subseteq &
  \EQ(\ff{P'}_\rho)*r*s, \;\;\mbox{ and}
\\
  {[\EQ(\ff{Q}_\rho)*r*s \rightarrow \EQ(\Good)]} & \subseteq &
  {[\EQ(\ff{Q'}_\rho)*r*s \rightarrow \EQ(\Good)]}.
\end{array}
$$
Thus, $h_0[\EQ(\ff{P'}_\rho)*r*s]h_1$ and
$k_0[\EQ(\ff{Q'}_\rho)*r*s \rightarrow \EQ(\Good)]k_1$.
These two relationships imply the required 
$\ff{M}_{\rho,\eta_0}(h_0,k_0)\bigl[\EQ(\Good)\bigr]
 \ff{M}_{\rho,\eta_1}(h_1,k_1)$, because
 $(\rho,\eta_0,\eta_1,r)$ satisfies $\mtri{P'}{M}{Q'}$.

The sixth case is the rule for introducing
existential quantification for assertions:
$$
(\forall i. \mtri{P}{M}{Q})
 \Rightarrow \mtri{\exists i.P}{M}{\exists i.Q}.
$$
Consider $(\rho,\eta_0,\eta_1,r)$ that satisfies
$\forall i. \mtri{P}{M}{Q}$.  We should show that 
$(\rho,\eta_0,\eta_1,r)$  
satisfies $\mtri{\exists i.P}{M}{\exists i.Q}$, i.e.,
$$
   \squad
     {\EQ(\ff{\exists i.P}_\rho)*r}
     {\ff{M}_{\rho,\eta_0}}
     {\ff{M}_{\rho,\eta_1}}
     {\EQ(\ff{\exists i.Q}_\rho)*r}.
$$
Pick arbitrary $h_0,h_1 \in \Heap$, $k_0,k_1 \in \cont$,
and $s \in\cR$ such that 
$$
   h_0[\EQ(\ff{\exists i.P}_\rho)*r*s]h_1
   \;\;\mbox{and}\;\;
   k_0[(\EQ(\ff{\exists i.Q}_\rho)*r*s) \rightarrow \EQ(\Good)]k_1.
$$
By the definition of $\EQ$, $\ff{\exists i.P}$ and 
$\ff{\exists i.Q}$, these two conjuncts imply the
existence of $v$ and $\rho'$ such that
$$
   \rho' = \rho[i\bind v],\;\;
   h_0[\EQ(\ff{P}_{\rho'})*r*s]h_1,\;\;
   \mbox{and}\;\;
   k_0[(\EQ(\ff{Q}_{\rho'})*r*s) \rightarrow \EQ(\Good)]k_1.
$$
From what we have just shown, we derive the conclusion 
as follows:
$$
\begin{array}{@{}rl@{}}
&
   (h_0[\EQ(\ff{P}_{\rho'})*r*s]h_1)
   \;\wedge\;
   (k_0[(\EQ(\ff{Q}_{\rho'})*r*s) \rightarrow \EQ(\Good)]k_1)
\\
\implies\;
&
   \ff{M}_{\rho',\eta_0}(h_0,k_0)[\EQ(\Good)]
   \ff{M}_{\rho',\eta_1}(h_1,k_1)
\;\,\hfill
   (\mbox{since }
    (\rho,\eta_0,\eta_1,r) \models \forall i.\mtri{P}{M}{Q})
\\
\implies\;
&
   \ff{M}_{\rho,\eta_0}(h_0,k_0)[\EQ(\Good)]
   \ff{M}_{\rho,\eta_1}(h_1,k_1)
\quad\hfill
   (\mbox{since } i \not\in \FV(M)).
\end{array}
$$

The seventh case is the disjunction rule. Suppose
that $(\rho,\eta_0,\eta_1,r)$ satisfies triples
$\mtri{P}{M}{Q}$ and $\mtri{P'}{M}{Q'}$. 
Consider
$h_0,h_1 \in \Heap$, $s \in \cR$,
and $k_0,k_1 \in \cont$, such that
$$
   h_0\bigl[\EQ(\ff{P\vee P'}_\rho)*r*s\bigr]h_1\;\wedge\;
   k_0\bigl[\EQ(\ff{Q \vee Q'}_\rho)*r*s \rightarrow 
             \EQ(\Good)\bigr]k_1.
$$
By the definition of $\EQ(\ff{P\vee P'}_\rho)$, 
heaps $h_0$ and $h_1$ are related by
$\EQ(\ff{P}_\rho)*r*s$ or $\EQ(\ff{P'}_\rho)*r*s$.
Without loss of generality, we assume that
\begin{equation}\label{eqn:disjunction-one}
h_0\bigl[\EQ(\ff{P}_\rho)*r*s\bigr]h_1.
\end{equation}
Since $\EQ$ is monotone and $*$ preserves the 
subset order for relations,
relation $\EQ(\ff{Q \vee Q'}_\rho)*r*s \rightarrow \EQ(\Good)$
is a subset of $\EQ(\ff{Q}_\rho)*r*s \rightarrow \EQ(\Good)$,
and so,
\begin{equation}\label{eqn:disjunction-two}
k_0\bigl[\EQ(\ff{Q}_\rho)*r*s \rightarrow \EQ(\Good)\bigr]k_1.
\end{equation}
By the supposition, $(\rho,\eta_0,\eta_1,r)$ satisfies
$\mtri{P}{M}{Q}$. Thus, the relationships $\ref{eqn:disjunction-one}$
and $\ref{eqn:disjunction-two}$ imply the required
$$
  \ff{M}_{\rho,\eta_0}(h_0,k_0)
  \bigl[\EQ(\Good)\bigr]
  \ff{M}_{\rho,\eta_1}(h_1,k_1).
$$

The eighth case is the rule for conditional statement.
Suppose that $(\rho,\eta_0,\eta_1,r)$ satisfies 
$\mtri{P \wedge E=F}{M}{Q}$ and 
$\mtri{P \wedge E\not=F}{N}{Q}$. Consider
$h_0,h_1 \in \Heap$, $s \in\cR$,
and $k_0,k_1 \in \cont$, such that
$$
   h_0\bigl[\EQ(\ff{P}_\rho)*r*s\bigr]h_1\;\wedge\;
   k_0\bigl[\EQ(\ff{Q}_\rho)*r*s \rightarrow \EQ(\Good)\bigr]k_1.
$$
We do the case analysis depending on whether $\ff{E}_\rho = \ff{F}_\rho$.
Suppose that $\ff{E}_\rho = \ff{F}_\rho$. In this case,
$h_0\bigl[\EQ(\ff{P \wedge E=F}_\rho)*r*s\bigr]h_1$,
and
\begin{equation}\label{eqn:ifz}
   \ff{\ifz\,(E{=}F)\,\then\,M\,\melse\,N}_{\rho,\eta_j}(h_j,k_j) = 
   \ff{M}_{\rho,\eta_j}(h_j,k_j) \mbox{ for all $j=0,1$}.
\end{equation}
Using these facts, we derive the conclusion as follows:
$$
\begin{array}{@{}ll@{}}
&
   h_0\bigl[\EQ(\ff{P \wedge E=F}_\rho)*r*s\bigr]h_1 \;\wedge\;
   k_0\bigl[\EQ(\ff{Q}_\rho)*r*s \rightarrow \EQ(\Good)\bigr]k_1
\\
\implies &
   \ff{M}_{\rho,\eta_0}(h_0,k_0)
   \bigl[\EQ(\Good)\bigr]
   \ff{M}_{\rho,\eta_1}(h_1,k_1)
\\
\implies &
   \ff{\ifz\,(E{=}F)\,\then\,M\,\melse\,N}_{\rho,\eta_0}(h_0,k_0)  
   \bigl[\EQ(\Good)\bigr]
   \ff{\ifz\,(E{=}F)\,\then\,M\,\melse\,N}_{\rho,\eta_1}(h_1,k_1) .
\end{array}
$$
The first implication follows from our assumption that 
$\mtri{P \wedge E=F}{M}{Q}$ is satisfied by
$(\rho,\eta_0,\eta_1,r)$, 
and the second follows from the equation
\ref{eqn:ifz} above. The other case $\ff{E}_\rho \not= \ff{F}_\rho$
can be proved similarly, so it is omitted here.

The ninth case is the rule for sequential composition.
Suppose that $(\rho,\eta_0,\eta_1,r)$ satisfies
$\mtri{P}{M}{P_0}$ and $\mtri{P_0}{N}{Q}$. Consider
$h_0,h_1 \in \Heap$, $s \in\cR$,
and $k_0,k_1 \in \cont$, such that
$$
   h_0\bigl[\EQ(\ff{P}_\rho)*r*s\bigr]h_1\;\wedge\;
   k_0\bigl[\EQ(\ff{Q}_\rho)*r*s \rightarrow \EQ(\Good)\bigr]k_1.
$$
Let $k'_j$ be $\lambda h'_j.\ff{N}_{\rho,\eta_j}(h'_j,k_j)$.
Since $(\rho,\eta_0,\eta_1,r) \models \mtri{P_0}{N}{Q}$,
$$
  k'_0 = \lambda h'_0.\ff{N}_{\rho,\eta_0}(h'_0,k_0)
  \bigl[\EQ(\ff{P_0}_\rho)*r*s \rightarrow \EQ(\Good)\bigr]
  \lambda h'_1.\ff{N}_{\rho,\eta_1}(h'_1,k_1) = k'_1.
$$
Since $(\rho,\eta_0,\eta_1,r) \models \mtri{P}{M}{Q_0}$,
the above relationship between $k'_0$ and $k'_1$ implies
$$
\ff{M}_{\rho,\eta_0}(h_0,k'_0)
  \bigl[\EQ(\Good)\bigr]
  \ff{M}_{\rho,\eta_1}(h_1,k'_1).
$$ 
This
gives the conclusion, because
$\ff{M;N}_{\rho,\eta_j}(h_j,k_j)$ is
equal to
$\ff{M}_{\rho,\eta_j}(h_j,k'_j)$, for all $j=0,1$.

The last case is the rule for fixed point induction. We note
two properties of $C$ and $\gamma$.
\begin{enumerate}
\item For all $\rho,\eta$, if $\eta(x) = \bot$, then
    $\ff{C(x)}_{\rho,\eta} = \bot$.
\item For all $(\rho,\eta_0,\eta_1,r)$, the following set is admissible:
$$
  \{(m_0,m_1) \mid (\rho,\eta_0[x\bind m_0],\eta_1[x\bind m_1],r) \models \gamma(x)\}.
$$
\end{enumerate}
These properties can be proved by a straightforward induction on the
structure of $C$ and $\gamma$. The soundness of the induction rule 
follows from the second property.
\qed

\section{A General Construction}
\label{sec:general-construction}
Our Kripke semantics of specifications presented in the previous section is
in fact an instance of a general, abstract construction that allows one to
interpret a specification logic with
higher-order frame rules. In this section, we describe the general
construction. The remaining part of the paper can be read and understood
without reading this section, in which we assume some basic knowledge
of categorical logic (see, e.g.,~\cite{biering-birkedal-torpsmith-esop05}
for a quick recap).

Before explaining our construction, we remind the reader of FM-cousins of
monoid, preorder, Heyting algebra and complete Heyting algebra. For FM-sets
$A,B,C$, we call an element $a_0 \in A$, a function $f : A \times B
\rightarrow C$ or a relation $r \subseteq A\times B$ {\em equivariant\/}
when they preserve the permutation action in the following sense: for all
$a \in A$, $b\in B$ and $\pi \in \perm$,
$$
   (\pi \cdot a_0 = a_0)
   \;\;\wedge\;\;
   (\pi \cdot (f(a,b)) = f(\pi \cdot a, \pi \cdot b)) 
   \;\;\wedge\;\;
   ((a,b) \in r \iff (\pi \cdot a, \pi \cdot b) \in r).
$$
An FM-monoid is an FM-set $M$ with monoid operations
$(I \in M,\;\; *:M\times M \rightarrow M)$ such that
$I$ and $*$ are equivariant,
and an FM-preorder is an FM-set $A$ with an equivariant
preorder $\sqsubseteq$ on $A$. An FM-Heyting algebra
is an FM-poset $(A,\sqsubseteq)$ with operations
$$
   \bot,\top \in A,\;\;\;\;\mbox{and}\;\;\;\;
   \sqcup,\sqcap,\Rightarrow : A \times A \rightarrow A,
$$
such that all of those operations are equivariant and
$(A,\sqsubseteq,\bot,\top,\sqcup,\sqcap)$ forms a Heyting algebra.
Finally, an FM-complete Heyting algebra is
an FM-Heyting algebra
$(A,\sqsubseteq,\bot,\top,\sqcup,\sqcap,\\ \Rightarrow)$ such 
that every finitely supported subset of $A$ has a least
upper bound and a greatest lower bound.

Our construction starts with an FM-monoid $(M,I,*)$ in which
$*$ is commutative.
The FM-monoid $(M,I,*)$ generalizes the set of
finitely supported binary relations $r$ on heaps,
where the monoid unit $I$ 
is the singleton relation ${([],[])}$ of 
two empty heaps and the monoid operator $*$ 
is the separating conjunction
for relations. Intuitively, each $m$ in $M$ represents 
information about the internal resource invariants of modules,
and the $*$ operator of $M$
is used to combine two pieces
of information that describe disjoint resources. 
Throughout this section, we assume given
a fixed FM-monoid $(M,I,*)$ with $*$ commutative,
and describe a construction over this FM-monoid.

First, we define a preorder $\sqsubseteq$ for $M$:
$$
  m \sqsubseteq n \iff
  \exists m'. m*m' = n.
$$
Intuitively, $m\sqsubseteq n$ means that $n$ is an extension
of $m$ with information about additional disjoint resources. 
\begin{lem}
$(M,\sqsubseteq)$ is an FM-preorder.
\end{lem}

It is well known that Kripke models of intuitionistic propositional
logic are obtained by taking the upwards closed subsets of a preorder and
that the upwards closed subsets form a complete Heyting algebra. 
Thus, our next step is to form such a model over 
$M$, but in the world of FM-sets.
Hence we construct an FM-complete Heyting algebra
$L(M)$ whose underlying set $L$ consists
of finitely supported upwards closed subsets of $M$,
and which is ordered by subset inclusion,
denoted $\sqsubseteq_L$. The
Heyting operations for $L$ are defined in the standard way:
when $M_0,M_1 \in L(M)$,
$$
\begin{array}{r@{\;}c@{\;}l@{\qquad\qquad}r@{\;}c@{\;}l}
\bot & \defeq & \emptyset 
&
\top & \defeq & M
\\
M_0 \sqcup M_1 & \defeq & M_0 \cup M_1
&
M_0 \sqcap M_1 & \defeq & M_0 \cap M_1
\\
\multicolumn{6}{c}{
M_0 \Rightarrow M_1 \;\defeq\; 
\{m \,\mid\, \forall m'.\; (m \sqsubseteq m' \wedge m' \in M_0)
\implies m' \in M_1\}.
}
\end{array}
$$
\begin{lem}
$(L(M),\sqsubseteq_L,\bot,\top,\sqcup,\sqcap,\Rightarrow)$
is an FM-complete Heyting algebra.
\end{lem}

The lattice $L(M)$ has two interesting properties, which
we used in our semantics of separation logic.
The first property is that the $\Rightarrow$ operator
involves quantification over information about disjoint
resources:
\begin{lem}
An element $m$ belongs to $M_0\Rightarrow M_1$ if and only if
$$
\forall m'.\; m*m' \in M_0 \implies m*m' \in M_1.
$$
\end{lem}
The second property is about the operator that frames in
information about disjoint resources. We define
a binary operator $-\otimes - : L(M) \times M \rightarrow L(M)$ 
by
$$
    M_0 \otimes m \;\;\defeq\;\; 
   \{ m' \,\mid\, m' * m \in M_0\}.
$$
\begin{lem}
The function $-\otimes -$ is well-defined, and it satisfies
the following three properties:
\begin{enumerate}
\item $-\otimes m$ 
commutes with $\Rightarrow$ and all the existing
least upper bounds or greatest lower bounds of subsets of $L(M)$.
\item $(M_0 \otimes m) \otimes m' = M_0 \otimes (m*m')$ 
for all $M_0 \in L(M)$ and $m,m' \in M$. 
\item  $M_0$ is a subset of $M_0 \otimes m$, 
for every $M_0 \in L(M)$.
\end{enumerate}
\end{lem}
In our semantics of separation logic, we used this $\otimes$ operator
to interpret invariant extension $\varphi\otimes P$, and designed
its proof rules, based on the general 
properties of $\otimes$ summarized in the above lemma.

Finally, we construct a hyperdoctrine $\mathsf{FMSet}(-,L(M))$, which can
be used to interpret the specification logic, including quantifiers and
invariant extensions (i.e., $\varphi \otimes P$).
\begin{lem}
\label{lem:hyperdoctrine}
$\mathsf{FMSet}(-,L(M))$ satisfies
all the axioms for hyperdoctrines, thereby allowing
the interpretation of intuitionistic predicate logic.
\end{lem}
For each $m \in M$,
consider the fibred endo-functor 
$$
\begin{array}{rcl}
\mathsf{FMSet}(-,-\otimes m)
& : & 
\mathsf{FMSet}(-,L(M))
\rightarrow 
\mathsf{FMSet}(-,L(M)),
\end{array}
$$
which maps a predicate $\varphi$ over $X$,
that is, an equivariant function $\varphi$ from
$X$ to $L(M)$, to $(-\otimes m) \circ \varphi$.
\begin{lem}
The fibred functor $\mathsf{FMSet}(-,-\otimes m)$
preserves $\bot,\top,\sqcup,\sqcap,\Rightarrow$ in each fibre
and commutes with quantifiers $\exists$ and $\forall$.
\end{lem}

In summary, the previous two lemmas provide alternative proofs to large parts
of Lemma~\ref{lemma:axiom-extension} and Theorem~\ref{thm:soundness}. In
the proof of the latter theorem in the previous section we omitted the
detailed proof of soundness of the rules for predicate logic; it is a
consequence of the above Lemma~\ref{lem:hyperdoctrine}.  Finally, we remark
that the general construction actually gives us more than we use in the
previous section: First, since we have a hyperdoctrine, we in fact have a
model of \emph{higher-order} specification logic in which one can also
model quantification over specifications.  Second, $L(M)$ is in fact not
only an FM-complete Heyting algebra but an FM-complete BI algebra. This
means that we can have $*$ and $-\!\!*$ connectives also for
specifications.  We have not yet made use of these additional facts.

\section{Properties of Semantic Quadruples}
\label{sec:quadruples}

In this section, we prove two properties of semantic quadruples.
The first clarifies the connection between our new interpretation 
of Hoare triples and the standard interpretation, 
and the second shows how our cps-style 
semantic quadruples are related to a more direct way
of relating two commands.

First, we consider the relation between semantic quadruples and
Hoare triples.
Define an operator $\cps$ that cps-transforms a state transformer
semantically: 
$$
\begin{array}{rcl}
   \cps & : &
   (\Heap \rightarrow (\Heap + \{\error\})_\bot) 
   \;\rightarrow\;
   (\Heap \times \cont \rightarrow \Obs)
\\
   \cps(c) & \defeq &
   \lambda (h,k).\;
   \ifthenElse
      {(c(h)\not\in \{\bot,\error\})}
      {k(c(h))}
      {c(h)}.
\end{array}
$$
\begin{prop}\label{prop:triple-quadruple}
For all $p,q \subseteq \Heap$ and all 
$c \in \Heap \rightarrow (\Heap + \{\error\})_\bot$,
the quadruple $\squad{\EQ(p)}{\cps(c)}{\cps(c)}{\EQ(q)}$ holds
iff the two conditions below hold:
\begin{enumerate}
\item for every $h$ in $p$,
         either $c(h) = \bot$ or $c(h) \in q$, hence
         $c(h)$ cannot be $\error$;
\item for every $h$ in $p$ and $h_1$ such that $h\# h_1$,
   \begin{enumerate}
   \item if $c(h) = \bot$, then $c(h\bullet h_1) = \bot$,
   \item if $c(h) \not=\bot$, then $c(h)\bullet h_1$ is defined and equal
         to $c(h\bullet h_1)$.
   \end{enumerate}
\end{enumerate}
\end{prop}
Note that the first condition is the usual meaning
of Hoare triples, and the second is the locality
condition of commands in separation logic restricted to 
heaps in $p$ \cite{reynolds02}. Since the locality condition merely
expresses the parametricity of commands with respect to new heaps, 
the proposition indicates that our interpretation of triples is the usual one
enhanced by an additional parametricity requirement.

\proof[Proof of Proposition~\ref{prop:triple-quadruple}]
$(\Rightarrow)$ Pick an arbitrary heap $h$ in $p$. Let $k$ be the
continuation defined by 
$$
   k(h) \defeq \ifthenElse{(h\in q)}{\bot}{\error}. 
$$
Then, $k[\EQ(q) \rightarrow \EQ(\Good)]k$ and $h[\EQ(p)]h$. By
the assumption on the validity of the quadruple,
$\cps(c)(h,k)[\EQ(\Good)]\cps(c)(h,k)$. By the definition of $k$,
this relationship on $\cps(c)$ implies that $\cps(c)(h,k) = \bot$,
which in turn gives 
$$
(c(h) = \bot) \vee (c(h) \in \Heap \wedge k(c(h)) = \bot).
$$ 
The second disjunct of this disjunction is equivalent to $c(h) \in q$
because $k(h') = \bot \iff h' \in q$. So, the
disjunction gives the first condition.

For the second condition, consider $h,h_1$ such that 
$h \in p$ and $h \# h_1$. Let $r$ be the relation
$\{([],h_1)\}$ on heaps, and define three
continuations $k_0,k_1,k_2$ as follows:
$$
\begin{array}{rcl}
  k_0(h') & \defeq & \normal,
\\
  k_1(h') & \defeq & \ifthenElse{(h'=c(h))}{\normal}{\bot},
\\
  k_2(h') & \defeq & \ifthenElse{(c(h) \in \Heap \,\wedge\, h'=c(h)\bullet h_1)}{\normal}{\bot}.
\end{array}
$$
By the definition of $r$ and $k_i$, we have that
$$
   h[\EQ(p)*r](h\bullet h_1),\;\;
   k_0[\EQ(q)*r \rightarrow \EQ(\Good)]k_0,\;\;
   \mbox{and}\;\;
   k_1[\EQ(q)*r \rightarrow \EQ(\Good)]k_2.
$$
To see why the third relationship holds, note that if 
$h'_1[\EQ(q)*r]h'_2$, then $h'_1\bullet h_1$ is defined and
$h'_2 = h'_1 \bullet  h_1$. Thus, $h'_1 = c(h)$ holds precisely 
when $c(h) \in \Heap \,\wedge\, h'_2 = c(h)\bullet h_1$ holds. 
This implies that $k_1(h'_1) = \normal$ iff $k_2(h'_2) = \normal$.
Now, by the assumption on
the validity of the quadruple, we have that
$$
   \cps(c)(h,k_0)[\EQ(\Good)]\cps(c)(h\bullet h_1,k_0)
   \;\;\mbox{and}\;\;
   \cps(c)(h,k_1)[\EQ(\Good)]\cps(c)(h\bullet h_1,k_2).
$$
The first conjunct about $k_0$ implies that if
$c(h) = \bot$, then $c(h\bullet h_1) = \bot$, and the
second conjunct about $k_1,k_2$ implies that if
$c(h) \not= \bot$, then $c(h\bullet h_1) = c(h)\bullet h_1$.

$(\Leftarrow)$ Consider a relation $r$ on heaps and
pick heaps $h_1,h_2$ and continuations $k_1,k_2$ such that
$$
   h_1[\EQ(p)*r]h_2 \;\;\mbox{and}\;\;
   k_1[\EQ(q)*r \rightarrow \EQ(\Good)]k_2.
$$
Then, there exist two splittings $h'_1\bullet h''_1 = h_1$ and
$h'_2\bullet h''_2 = h_2$ such that $h'_1 = h'_2 \in p$ and
$h''_1[r]h''_2$. If $c(h_1) = \bot$, then $c(h'_1) = \bot$
by the condition (2-b) of the assumption, and 
$c(h'_2) = \bot$ by the condition (2-a) of the assumption.
Thus, in this case, we have 
$\cps(c)(h_1,k_1) = \cps(c)(h_2,k_2) = \bot$ and
$\cps(c)(h_1,k_1)[\EQ(\Good)]\cps(c)(h_2,k_2)$, as desired.
Otherwise, i.e., if $c(h_1) \not= \bot$, then $c(h'_1) \not=
\bot$ by the condition (2-a). Thus, by the condition (2-b),
we have that $c(h_1)  = c(h'_1) \bullet  h''_1$ and
$c(h_2)  = c(h'_1) \bullet  h''_2$. Since $c(h'_1) \in q$ by
the condition (1),  
$$
   c(h_1) = c(h'_1)\bullet h''_1[\EQ(q)*r]c(h'_1)\bullet h''_2 = c(h_2).
$$
This implies $\cps(c)(h_1,k_1)[\EQ(\Good)]\cps(c)(h_2,k_2)$,
as desired.
\qed

Next, we relate our cps-style notion of semantic quadruples 
to the direct-style alternative. The notion underlying 
this relationship is the observation closure, denoted $(-)^\mperp$.
For each FM-cpo $D$ and relation 
$r \subseteq D \times D$, we define two relations, 
$r^\bot$ on $[D \rightarrow \Obs]$ and $r^\mperp$ on $D$,
as follows:
$$
\begin{array}{rcl}
   k_1[r^\bot]k_2 
   & \defliff &
   \forall d_1,d_2 \in D.\;
       (d_1[r]d_2 \implies k_1(d_1)[\EQ(\Good)]k_2(d_2)),
\\
   d_1[r^\mperp]d_2 
   & \defliff &
   \forall k_1,k_2 \in [D \rightarrow \Obs].\;
       (k_1[r^\bot]k_2 \implies k_1(d_1)[\EQ(\Good)]k_2(d_2)). 
\end{array}
$$
Operator $(-)^\bot$ dualizes a relation on $D$ to
one on observations on $D$, and $(-)^\mperp$ closes a given 
relation $r$ under observations. 
\begin{prop}\label{prop:quadruple-direct}
Let $r,s$ be relations in $\cR$.
Consider functions $c_1,c_2$ from 
$\Heap$ to $(\Heap  + \{\error\})_\bot$.
A quadruple $\squad{r}{\cps(c_1)}{\cps(c_2)}{s}$ holds, iff
$$
\forall (r',h_1,h_2).\;
h_1[r*r']h_2 
\implies
(c_1(h_1){=}c_2(h_2){=}\bot \vee c_1(h_1)[(s*r')^\mperp]c_2(h_2)).
$$
\end{prop}
This proposition shows that our semantic quadruples are close
to what one might expect at first for relating two commands
parametrically. The only difference is that our
quadruple always closes the post-relation $s*r'$ under 
observations.

\proof[Proof of Proposition~\ref{prop:quadruple-direct}]
$(\Rightarrow)$
Consider $r',h_1,h_2$ such that $h_1[r*r']h_2$. We
first show that 
$$
  c_1(h_1) = \bot \iff c_2(h_2) = \bot.
$$
 Let $k$ be the continuation $\lambda h'.\normal$.
Then, $k[s*r' \rightarrow \EQ(\Good)]k$. By the assumption
on the quadruple for $\cps(c_1),\cps(c_2)$, we have that
$$
   \cps(c_1)(h_1,k)[\EQ(\Good)]\cps(c_2)(h_2,k).
$$ 
This relationship implies that
$c_1(h_1) = \bot \iff c_2(h_2) = \bot$, because 
$c_i(h_i) = \bot \iff \cps(c_i)(h_i,k) = \bot$ by
the choice of $k$.

Next, we prove that if $c_1(h_1)\not=\bot$ or 
$c_2(h_2)\not=\bot$, then $c_1(h_1)[(s*r')^\mperp]c_2(h_2)$.
By what we have just shown, $c_1(h_1)\not=\bot$ iff
$c_2(h_2)\not=\bot$. We will assume that neither $c_1(h_1)$
nor $c_2(h_2)$ is $\bot$. Take two continuations $k_1,k_2$
such that $k_1[(s*r')^\bot]k_2$, i.e.,
$k_1[s*r' \rightarrow \EQ(\Good)]k_2$. Since the quadruple
$\squad{r}{\cps(c_1)}{\cps(c_2)}{s}$ holds by assumption
and $h_1[r*r']h_2$, we have that 
$$
  \cps(c_1)(h_1,k_1)[\EQ(\Good)]\cps(c_2)(h_2,k_2).
$$
Since both $c_1(h_1)$ and $c_2(h_2)$ are different from $\bot$,
 the relationship is equivalent to
$$
k_1(c_1(h_1))[\EQ(\Good)]k_2(c_2(h_2)).
$$ 
We have just shown
that $c_1(h_1)[(s*r')^\mperp]c_2(h_2)$.

$(\Leftarrow)$ Pick an arbitrary relation $r'$, heaps $h_1,h_2$ and 
continuations $k_1,k_2$ such that $h_1[r*r']h_2$ and 
$k_1[s*r'\rightarrow \EQ(\Good)]k_2$ (i.e.,
$k_1[(s*r')^\bot]k_2$.) By the assumption of this if direction,
either  $c_1(h_1) = c_2(h_2) = \bot$ or 
$c_1(h_1)[(s*r')^\mperp]c_2(h_2)$.
In the first case,
$$
\cps(c_1)(h_1,k_1) = \bot\,[\EQ(\Good)]\,\bot = \cps(c_2)(h_2,k_2),
$$
and in the second case, both $c_1(h_1)$ and $c_2(h_2)$ are in $\Heap$,
so that
$$
  \cps(c_1)(h_1,k_1) = k_1(c_1(h_1))
  \,[\EQ(\Good)]\,
  k_2(c_2(h_2)) = \cps(c_2)(h_2,k_2).
$$
The conclusion follows from these two relationships.
\qed


\section{Abstraction Theorem}
\label{sec:abstraction-theorem}

The abstraction theorem below formalizes that well-specified programs
(specified in separation logic with implicit quantification over internal
resource invariants by frame rules) behave relationally parametrically in
internal resource invariants. The easiest way to understand this intuition 
may be from the corollary following the theorem.

Some readers might feel that it
is too much to call the abstraction theorem a ``theorem'' since 
it really is a trivial corollary of the soundness theorem
--- but that is just as it should be:
the semantics was defined to achieve that. 

\begin{thm}[Abstraction Theorem]\label{thm:abstraction}
If $\Delta \mid \Gamma \,\vdash\, \varphi$ is provable in
the logic, then for all 
$(\rho,\eta_0,\eta_1,r) \in \ff{\Delta} \times \ff{\Gamma}^2 \times \cR$,
we have that $(\rho,\eta_0,\eta_1,r) \models \varphi$.
\end{thm}
\proof
  By Theorem~\ref{thm:soundness}, we get that 
  $\Delta \mid \Gamma \,\vdash\, \varphi$ is
  valid, which is just what the conclusion expresses.
\qed

\begin{cor}\label{cor:simple-abstraction}
Suppose that  $\Delta \mid x\colon \com \,\vdash\, 
     \mtri{P_1}{x}{Q_1} \Rightarrow \mtri{P}{M}{Q}$
is provable in the logic. Then for all 
$(\rho,c_0,c_1,r)$,
if \/ $\squad{\EQ(\ff{P_1}_\rho)*r}{c_0}{c_1}{\EQ(\ff{Q_1}_\rho) * r}$
holds, then
$$
\squad
     {\EQ(\ff{P}_\rho)*r}
     {\ff{M}_{[x\bind c_0]}}
     {\ff{M}_{[x\bind c_1]}}
     {\EQ(\ff{Q}_\rho) * r}
$$ 
holds as well.
\end{cor}
Intuitively, $x$ corresponds to a module with a single 
operation, and $M$ a client of the module. This corollary
says that if we prove a property of the client $M$, assuming
only an abstract external specification $\mtri{P_1}{x}{Q_1}$ of
the module, the client cannot tell apart two different 
implementations $c_0,c_1$ of the module, as long as
$c_0,c_1$ have identical external behavior. The 
four instances of $\EQ$ in the proposition formalize that
the external behaviors of $c_0,c_1$ are identical
and that the client $M$ behaves the same externally
regardless of whether it is used with $c_0$ or $c_1$. 
The relation $r$ is a simulation relation for
internal resource invariants of $c_0$ and $c_1$.
\proof[Proof of Corollary~\ref{cor:simple-abstraction}]
Define environments $\eta_0,\eta_1$ and heap sets $p,p_1,q,q_1$ as follows:
$$
\eta_0 = [x\bind c_0],\; \eta_1 = [x\bind c_1],\;\mbox{and}\;
(p_1,q_1,p,q) = (\ff{P_1}_\rho,\ff{Q_1}_\rho,\ff{P}_\rho,\ff{Q}_\rho).
$$
By Theorem~\ref{thm:abstraction}, we have, for any $r$, that
$
   (\rho,\eta_0,\eta_1,r) \models \mtri{P_1}{x}{Q_1} \Rightarrow \mtri{P}{M}{Q}.
$
From this, we derive the conclusion of the proposition:
$$
\begin{array}{@{}cl@{}}
& (\rho,\eta_0,\eta_1,r) \models \mtri{P_1}{x}{Q_1} \Rightarrow \mtri{P}{M}{Q}
\\
{\implies} 
& (\forall s \in \cR.\;
   (\rho,\eta_0,\eta_1,r*s) \models \mtri{P_1}{x}{Q_1} \implies
   (\rho,\eta_0,\eta_1,r*s) \models \mtri{P}{M}{Q}) 
\\
{\implies} 
& ((\rho,\eta_0,\eta_1,r) \models \mtri{P_1}{x}{Q_1} \implies
   (\rho,\eta_0,\eta_1,r) \models \mtri{P}{M}{Q}) 
\\
{\implies} 
& (\squad{\EQ(p_1)*r}{c_0}{c_1}{\EQ(q_1)*r} \implies
   \squad{\EQ(p)*r}{\ff{M}_{\eta_0}}{\ff{M}_{\eta_1}}{\EQ(q)*r}).
\end{array} 
$$
\qed

\section{Examples}
\label{sec:examples}

Our first example is the two implementations of a counter in
the introduction and the simple client $(\inc;\mread)$ in 
Example~\ref{exa:client-counter}. We remind the reader of
the implementations and the specification of the client in 
Figure~\ref{fig:exa-counter} (here we use the formally
correct 0 and 1 for the
fields named $\data$ and $\mnext$ in the introduction for
readability).
The figure also shows the concrete specifications of the
implementations. Note that the concrete specifications
describe that both implementations
use an internal cell $c.0$ to keep the value of the counter,
and that the second implementation stores the negated value
of the counter in this internal cell.

\begin{figure*}[t]
\hrule
$$
\begin{array}{rcl}
\multicolumn{3}{c}{
\begin{array}{@{}r@{}c@{}l@{}}
    \forall j.\; 
    \{\exists i.\,c\pointsto i,\blank * i\pointsto \blank,j\}
    & \inc_0 &
   \{\exists i.\,c\pointsto i,\blank * i\pointsto \blank,(j{+}1)\} 
\\
   \forall j.\;
   \{\exists i.\,c\pointsto i,\blank * i\pointsto \blank,j * g \pointsto \blank\}
    & \mread_0 &
    \{\exists i.\,c\pointsto i,\blank * i \pointsto \blank,j * g \pointsto \blank,j\}
\end{array}
}
\\[2.5ex]
    \inc_0 & \equiv & \mletin{i{=}c.0}{(\mletin{j{=}i.1}{i.1 := j{+}1})}
\\
    \mread_0 & \equiv & \mletin{i{=}c.0}{(\mletin{j{=}i.1}{g.1 := j})}
\\
\\
\multicolumn{3}{c}{
\begin{array}{@{}r@{}c@{}l@{}}
   \forall j.\; 
    \{\exists i.\,c\pointsto i,\blank * i\pointsto \blank,j\}
    & \inc_1 &
   \{\exists i.\,c\pointsto i,\blank * i\pointsto \blank,(j{-}1)\} 
\\
   \forall j.\; 
    \{\exists i.\,c\pointsto i,\blank * i\pointsto \blank,j * g \pointsto \blank\}
    & \mread_1 &
    \{\exists i.\,c\pointsto i,\blank * i \pointsto \blank,j * g \pointsto \blank,({-}j)\}
\end{array}
}
\\[2.5ex]
    \inc_1 & \equiv & \mletin{i{=}c.0}{(\mletin{j{=}i.1}{i.1 := j{-}1})}
\\
    \mread_1 & \equiv & \mletin{i{=}c.0}{(\mletin{j{=}i.1}{g.1 := {-}j})}
\\
\\
\multicolumn{3}{c}{
     \Delta \,\mid\, \Gamma \;\;\vdash\;\; 
     (\mtri{\emp}{\inc}{\emp}\land
      \mtri{g\pointsto\blank}{\mread}{g\pointsto\blank})
     \;\Rightarrow\; 
     \mtri{g\pointsto\blank}{\inc;\mread}{g\pointsto \blank}
}
\\
\multicolumn{3}{r}{
  \hfill(\mbox{where $\Delta \equiv \{g,c\}$ and
  $\Gamma \equiv \{\inc\colon \com, 
              \mread \colon\com\}$})
}
\end{array}
$$
\hrule
\caption{Two Implementations of a Counter and a Simple Client}
\label{fig:exa-counter}
\end{figure*}

Pick a location $l \in \Loc$ and an environment $\rho \in \ff{\{c,g\}}$ 
with $\rho(c) = l$, and define $f_0,f_1,g_0,g_1,b_0,b_1$ as follows:
$$
f_i \defeq \ff{\inc_i}_{\rho,[]},\;\;\; 
g_i \defeq \ff{\mread_i}_{\rho,[]},\;\;\;
b_i \defeq \ff{\inc;\mread}_{\rho,[\inc\bind f_i, \mread\bind g_i]}.
$$
Now, by the Abstraction Theorem, we get that, for all $r$,
\begin{equation}\label{eq:ex1}
\begin{array}{c}
  \bigl(
    \squad
    {\EQ(\ff{\emp}_\rho)*r}
    {f_0}
    {f_1}
    {\EQ(\ff{\emp}_\rho)*r}
    \;\land\;
    \squad
    {\EQ(\ff{g\pointsto\blank}_\rho)*r}
    {g_0}
    {g_1}
    {\EQ(\ff{g\pointsto\blank}_\rho)*r}\big) 
\\
 {}
 \Rightarrow
 {}
\\
  \squad
    {\EQ(\ff{g\pointsto \blank}_\rho)*r}
    {b_0}
    {b_1}
    {\EQ(\ff{g\pointsto \blank}_\rho)*r}.
\end{array}
\end{equation}
We now sketch a consequence of this result; for brevity we allow ourselves
to be a bit informal. Let $r$ be the following simulation relation
between the two implementations:
$$
  r \;\defeq\; \{\, (h_0,h_1)\,\mid\,
         \begin{array}[t]{@{}l@{}}
         \exists i \in \Loc.\,
         \exists n \in \sint.\,
         \exists v_0,v_1,v_0',v_1' \in \sval.\; \\
         \quad
          i \not= l \,\land\,
          h_0 = [c\bind i,v_0] \bullet [i\bind v_0',n]  \,\land\,
          h_1 = [c\bind i,v_1]\bullet [i\bind v_1',{-}n] \,\}.
        \end{array}
$$
Then one can verify that the antecedent of the implication
in~(\ref{eq:ex1}) holds, and thus conclude that 
$$
  \squad
    {\EQ(\ff{g\pointsto \blank}_\rho)*r}
    {b_0}
    {b_1}
    {\EQ(\ff{g\pointsto \blank}_\rho)*r}
$$
holds.  Take $(h_0,h_1)\in \EQ(\ff{g\pointsto
  \blank}_\rho)*r$, and denote the result of running $b_0$ on
$h_0$ by $h'_0$, and the result of running $b_1$ on $h_1$ by $h'_1$. We
then conclude that $h'_0$ will be of the form $h'_{00}\bullet h'_{01}$ and
that $h'_1$ will be of the form $h'_{10}\bullet h'_{11}$ with
$(h'_{01},h'_{11})\in r$ and with $(h'_{00}, h'_{10})\in\EQ(\ff{g\pointsto
  \blank}_\rho)$.

Thus the relation between the internal resource invariants is
maintained and, for the visible part, $b_0$ and $b_1$ both produce the
\emph{same heap} with exactly one cell.

\newcommand{\putf}{\mathsf{put}}
\newcommand{\getf}{\mathsf{get}}

The next example is a buffer of
size one, and it illustrates the ownership
transfer. Our buffer has operations $\putf$ and $\getf$.
Intuitively, $\putf(i)$ stores the value found at $i$ in the buffer,
and $\getf(j)$ retrieves the value stored in the buffer and stores
it at $j$.
We assume the following abstract specifications of this mutable abstract
data type:
$$ 
(\forall i.\,\mtri{i\pointsto\blank}{\putf(i)}{\emp}) 
\;\;\;\mbox{and}\;\;\; 
(\forall j.\,\mtri{j\pointsto\blank}{\getf(j)}{j\pointsto\blank}.
$$

\begin{figure*}[t]
\hrule
$$
\begin{array}{rcl}
\multicolumn{3}{c}{
\begin{array}{@{}r@{}c@{}l@{}}
    \forall i,v.\;
    \{i\pointsto \blank,v * k\pointsto \blank\}
    & {\putf_0(i)} &
    \{k\pointsto \blank,v\} 
\\
    \forall j,v.\;
    \{j\pointsto \blank * k\pointsto \blank,v\} 
    & {\getf_0(j)} &
    \{j\pointsto \blank,v * k\pointsto \blank,v\}
\end{array}
}
\\[2.5ex]
    \putf_0 & \equiv & \lambda i.\, \mletin{v=i.1}{(\free(i);k.1 := v)}
\\
    \getf_0 & \equiv & \lambda j.\, \mletin{v=k.1}{j.1 := v}
\\
\\
\multicolumn{3}{c}{
\begin{array}{@{}r@{}c@{}l@{}}
  \forall i,v.\;
  \{i\pointsto \blank,v * (\exists k'.\;k\pointsto k',\blank * k'\pointsto \blank)\}
  & \putf_1(i) &
  \{\exists k'.\; k\pointsto k',\blank * k'\pointsto \blank,v \}
\\
  \forall j,v.\;
  \{j\pointsto \blank * (\exists k'.\;k\pointsto k',\blank * k'\pointsto \blank,v)\}
 & \getf_1(j) &
  \{j\pointsto \blank,v * (\exists k'.\;k\pointsto k',\blank * k'\pointsto \blank,v) \}
\end{array}
}
\\[2.5ex]
    \putf_1 & \equiv & 
      \lambda i.\, \mletin{k' {=} k.0}{(\free(k'); k.0 {:=} i)}  
\\
    \getf_1 & \equiv &
     \lambda j.\,
        \mletin{k' {=} k.0}%
               {\mathsf{let}\; v {=} k'.1\;\mathsf{in}\; j.1 {:=} v}
\\
\\
\multicolumn{3}{c}{
     \Delta \,\mid\, \Gamma \;\;\vdash\;\; 
     (\forall i. \mtri{i\pointsto\blank}{\putf(i)}{\emp})\land
     (\forall j. \mtri{j\pointsto\blank}{\getf(j)}{j\pointsto\blank})
     \;\Rightarrow\; 
     \mtri{j\pointsto \blank}{c}{j\pointsto \blank}
}
\\
\multicolumn{3}{r}{
  \hfill(\mbox{where $\Delta \equiv \{j,k\}$ and
  $\Gamma \equiv \{\putf\colon \val \rightarrow \com, 
              \getf\colon \val \rightarrow \com\}$})
}
\\[1ex]
    c & \equiv &
  \mletin{i{=}\new}%
       {(i.1{:=} 5; \putf(i); \getf(j))}
\end{array}
$$
\hrule
\caption{Two Implementations of a Buffer and a Simple Client}
\label{fig:impl}
\end{figure*}

Figure~\ref{fig:impl} shows
two implementations of the buffer and a client,
as well as the concrete specifications for
the implementations and the specification for the client.
Note that the first implementation just uses one cell for the buffer
and that the implementation follows the intuitive description given above. 
The second implementation uses two cells for the buffer. The additional
cell is used to hold the cell pointed to by $i$ itself. Note that this
additional cell is transferred from the caller of $\putf_2(i)$, i.e.,
a client of the buffer. Finally, the specification of the client
describes the safety property of $c$, assuming the abstract specification 
for the buffer.

Pick $\rho \in \ff{\{j,k\}}$,
and define $f_0,f_1,g_0,g_1,c_0,c_1$ by
$$
f_i \defeq \ff{\putf_i}_{\rho,[]},\;\;\; 
g_i \defeq \ff{\getf_i}_{\rho,[]},\;\;\;
c_i \defeq \ff{c}_{\rho,[\putf\bind f_i, \getf\bind g_i]}.
$$
Our Abstraction Theorem gives that, for all $r$,
\begin{equation}\label{eq:ex2}
\begin{array}{l}
  \begin{array}[t]{@{}l@{}}
  (\forall v \in \sval.\;
  \squad
    {\EQ(\ff{i\pointsto \blank}_{\rho[i\bind v]})*r}
    {f_0(v)}
    {f_1(v)}
    {\EQ(\ff{\emp}_{\rho[i\bind v]})*r})
    \;\land\;
\\
  (\forall v \in \sval.\;
  \squad
    {\EQ(\ff{j\pointsto\blank}_{\rho[j\bind v]})*r}
    {g_0(v)}
    {g_1(v)}
    {\EQ(\ff{j\pointsto\blank}_{\rho[j\bind v]})*r})
  \end{array}
\\
 {}
 \;\;\Rightarrow\;\;
  \squad
    {\EQ(\ff{j\pointsto \blank}_\rho)*r}
    {c_0}
    {c_1}
    {\EQ(\ff{j\pointsto \blank}_\rho)*r}.
\end{array}
\end{equation}
This result implies that the client behaves the same no matter
whether we run it with the first or second implementation of the
buffer. To see this, let $l$ be $\rho(k)$ 
and define a simulation relation
$r$ between the two implementations:
$$
  r \;\defeq\; 
    \{\, (h_0,h_1)\,\mid\,
         \begin{array}[t]{@{}l@{}}
         \exists l' \in \Loc.\,\exists n,v_0,v_1,v_1'\in \sval.\; \\
          \qquad
          l \not= l' \;\land\;
          h_0 = [l\bind v_0,n] \;\land\;
          h_1 = [l\bind l',v_1]\bullet[l'\bind v_1',n]\, \}.
         \end{array}
$$
For this relation $r$, one can verify that the antecedent 
of the implication
in~(\ref{eq:ex2}) holds, and thus conclude that 
$$
  \squad
    {\EQ(\ff{j\pointsto \blank}_\rho)*r}
    {c_0}
    {c_1}
    {\EQ(\ff{j\pointsto \blank}_\rho)*r}
$$
holds. This quadruple says, in particular, 
that $c_0$ and $c_1$ map $\EQ(\ff{j\pointsto \blank}_\rho)*r$-related
heaps to $\EQ(\ff{j\pointsto \blank}_\rho)*r$-related heaps, 
which means
that they behave the same for cell $j$ and preserve the $r$ 
relation for the internal resource invariants of the two
implementations.


\section{Conclusion and Future Work}\label{sec:discussion-futurework}

We have succeeded in defining the first relationally parametric model of
separation logic. The model captures the informal idea that well-specified
clients of mutable abstract data types should behave parametrically in the
internal resource invariants of the abstract data type.

We see our work as a first step towards devising a logic for reasoning
about mutable abstract data types, similar in spirit to Abadi and
Plotkin's logic for parametricity~\cite{abadi-plotkin,BirkedalL:defp-mscs}. 
To this end, we also
expect to make use of the ideas of
relational separation logic in~\cite{yang-relational-separation-logic}
for reasoning about relations between different programs syntactically.
The logic should include a link between separation logic and
relational separation logic so that one could get a syntactic
representation of the semantic Abstraction Theorem and its corollary
presented above.

One can also think of our work as akin to the O'Hearn-Reynolds model for
idealized algol based on translation into a relationally parametric
polymorphic linear lambda calculus~\cite{ohearn-reynolds}.  In \emph{loc.
  cit.} O'Hearn and Reynolds show how to provide a better model of stack
variables for idealized algol by making a formal connection to
parametricity. Here we provide a better model for the more unwieldy world
of heap storage by making a formal connection to parametricity.

As mentioned in Section~\ref{sec:separation-logic}, 
the conjunction rule is not sound in our model. This is a consequence of
our interpretation, which ``bakes-in'' the frame rule by quantifying
over all relations $r'$. Indeed, using the characterization given by
Proposition~\ref{prop:quadruple-direct}, one sees that for the conjunction
rule 
\[\big(\squad{r_1}{\cps(c_1)}{\cps(c_2)}{s_1} \land
  \squad{r_2}{\cps(c_1)}{\cps(c_2)}{s_2} \big) \Longrightarrow
  \squad{r_1\land r_2}{\cps(c_1)}{\cps(c_2)}{s_1\land s_2}
\]
to hold, we would need something like $(r_1\land r_2)*r = (r_1*r)\land (r_2
*r)$ to hold. We ``bake-in'' the frame rule in order to get a model
that validates a wide range of higher-order frame rules and it is known
that already for second-order frame rules, the conjunction rule is not
sound without some restrictions on the predicates
involved~\cite{yang-ohearn-reynolds-popl04}. 
We don't know whether it is possible to
develop a parametric model in which the conjunction rule is sound. 

Future work further includes developing a parametric model for 
the higher-order version of separation logic with explicit quantification
over internal resource invariants. 
Finally, we hope that ideas similar to those presented here can be used to
develop parametric models for other recent approaches to mutable abstract
data types (e.g.,~\cite{
naumann-barnett}). 

\section*{Acknowledgments}
We would like to thank Nick Benton, Jacob Thamsborg and the anonymous
referees for their insightful comments. This work was supported by FUR
(FIRST). Yang was supported also by EPSRC.

\end{document}